\input amstex
\documentstyle{amsppt}
\magnification=1200
\def\cD{\Cal D}

\def\cN{\Cal N}
\def\cF{\Cal F}
\def\cB{\Cal B}
\def\cC{\Cal C}
\def\fW{\frak W}
\def\fH{\frak H}
\def\fM{\frak M}
\def\fD{\frak D}
\def\fN{\frak N}
\def\fA{\frak A}
\def\bR{\Bbb R}
\def\bC{\Bbb C}
\def\bN{\Bbb N}
\def\bE{\Bbb E}
\def\bI{\Bbb I}
\def\bL{\Bbb L}
\def\bl{\bold j}
\def\bk{\bold k}
\def\bT{\Bbb T}
\def\bZ{\Bbb Z}
\def\bh{\bold h}
\def\WA{\fW(\cD,\sigma)}
\def\GNS{({\fH}_{\omega},\pi_{\omega},\Omega_{\omega})}
\def\LDS{(\cD,\sigma,\{ T_{t} \})}
\def\GF{G(f_{1},\ldots,f_{n};t_{1},\ldots,t_{n})}
\def\EGF{G^{E}(f_{1},\ldots,f_{n};s_{1},\ldots,s_{n})}
\def\sk{\underline{s}^{k}}
\def\Wk{\underline{W}^{k}}
\def\sko{\underline{s}^{k \ast}}
\def\Wko{\underline{W}^{k \ast }}
\def\fk{\underline{f}^{k}}
\def\EGW{G^{E}(\Wk;\sk)}
\def\Wi{\underline{W}^{k-1}(i)}
\def\W#1{\underline{W}^{#1}}
\def\s#1{\underline{s}^{#1}}
\def\eo#1{{\frak E}^{\infty}_{#1}}
\def\eob#1{{\frak E}^{\beta}_{#1}}
\def\inprod#1#2{\langle #1, #2 \rangle}
\def\conj#1{\overline{#1}}
\def\pod#1{\underline{#1}}

\NoBlackBoxes
\topmatter
\title
Stochastically Positive Structures on Weyl Algebras.
The case of quasi-free states.
\endtitle
\rightheadtext{Stochastically Positive Structures on Weyl Agebras}
\author
R. Gielerak, L. Jak\'obczyk, R. Olkiewicz
\endauthor
\affil
Institute of Theoretical Physics, University of Wroc{\l}aw, Wroc{\l}aw,
Poland
\endaffil
\address
Institute of Theoretical Physics, University of Wroc{\l}aw,
Pl. M. Borna 9, 50-205 Wroc{\l}aw, Poland
\endaddress
\email
gielerak\@ift.uni.wroc.pl, ljak\@ift.uni.wroc.pl, rolek\@ift.uni.wroc.pl
\endemail
\thanks
Supported by KBN Grant 2PO3B 122 11
\endthanks
\abstract
We consider quasi-free stochastically positive ground and thermal states on 
Weyl algebras  in Euclidean time formulation. In particular, we obtain a new
derivation of a general form of thermal quasi-free state and give  
conditions when such state is stochastically positive i.e. when it defines
periodic stochastic process with respect to Euclidean time, so called 
thermal process. Then we show that thermal process completely determines 
modular structure canonically associated with quasi-free thermal state 
on Weyl algebra. We discuss a variety of examples connected with
free quantum field theories on globally hyperbolic stationary space-times and
models of quantum statistical mechanics.
\endabstract
\endtopmatter
\document
\head
I Introduction
\endhead
It is well known that the idea of analytic continuation to the imaginary 
time (Euclidean) variables is very fruitful for study several quantum
systems, especially Wightman quantum field theories on a flat space-time. 
From  general properties of Wightman distributions follows the 
possibility of analytic continuation to the Euclidean region, what gives 
an alternative, purely Euclidean, description in terms of so 
called Schwinger functions \cite{1, 2, 3, 4}. In the context of bosonic
fields,we obtain commutative objects, which are much easier to analyse 
\cite{5}. Particulary nice situation occurs in the case of free scalar
fields. As was noticed by Nelson \cite{6} (see also \cite{7}), the
corresponding Euclidean structure is given by a Gaussian Markov 
generalized random field. This observation and the general reconstruction 
theorem \cite{8}, gave powerful input for the development of the
constructive quantum field theory on a flat space-time (\cite{9, 10}
and references therein).
\par
One of the main objectives of the present paper is to develop 
systematically the rigorous Euclidean formalism in the context
of equilibrium states given by the KMS condition. As the interesting
application of general results, we consider Euclidean approach
to quantum field theories at non-zero temperature on globally
hyperbolic stationary space-times (\cite{11, 12}).
Some aspects of Euclidean formalism
in this context, were discussed  in the literature at various levels
of mathematical rigour (see e.g. \cite{13, 14, 15, 16}). But to
our knowledge, still there is the lack of a
systematic and mathematically rigorous approach
to these problems. In the context of quantum
statistical mechanics, Euclidean approach developed in 
\cite{17, 18, 19, 20, 21, 22, 23, 24, 25, 26, 27} also appears to be
very useful. In particular, for a class of quantum systems with
so called stochastic positivity, which are described by some
stochastic processes, methods of classical statistical mechanics
can be applied to study several questions like the existence and
properties of the limiting thermal states, phase transitions, etc.
(\cite{20, 21, 24, 25, 26, 27}).
\par
In the present paper  a general conceptual
framework for stochastically positive structures on Weyl algebras 
is introduced (Section II). As a starting point for further development,
we analyse in details the case of quasi-free states (Section III). Although 
basic structural results characterizing quasi-free thermal states have
been obtained long time ago \cite{28, 29, 30}, we present new (in our opinion
simpler) proof of the theorem giving general form of a quasi-free state
satisfying the KMS condition (see Theorem 3.1 and 3.7). Section 3.4
contains complete characterizations of stochastically positive quasi-free
KMS states and corresponding periodic stochastic processes. The important
problem of the equivalence of the modular structure given by a stochastic 
process (Theorem 3.3) with the canonical modular structure associated 
with a quasi-free KMS state is solved in full generality in Section 3.5. 
This result shows that in the case of quasi-free stochastically positive
states, all relevant informations about KMS structure are contained
in the commutative sector given by a thermal stochastic process.
Similar results are true also for the ground state case (Section 3.7).
Let us emphasize that it gives new possibilities for the description
of KMS structures in the case when interaction is present.
Having described quasi-free
systems in terms of stochastic processes, one may perturb them with
multiplicative-like functionals, thereby creating some new non-Gaussian
thermal process. Furthermore, given such a process, we can reproduce
its KMS structure. Some results into this direction were obtained in
\cite{24} for gentle perturbations of the free Bose gas in the
noncritical region of densities and in \cite{25} for the critical
case.
Essentially equivalent (via the Feynman-Kac formula)
type of perturbations which can be studied using this method
are central perturbations of a quasi-free state, by which we mean the
perturbations of the the form $"H_{\omega}+V"$, where $H_{\omega}$ is
the generator
of the free evolution in the thermal representation, and $V$ is an operator
affiliated with the von Neumann algebra $\pi_{\omega}(\frak A)^{\prime \prime}$,
for apropriate abelian subalgebra $\frak A$ of the Weyl algebra. That kind
of perturbations will be discussed in the second part of the present work.
Section IV contains several examples to which our general arguments
can be applied. In particular case of the scalar quantum field theory
on a globally hyperbolic stationary space-time, we give a general result
on the existence of Markov thermal process, which determines the
whole modular structure of the theory. Similar arguments are also valid
for ground state structure. In the case of a static space-time, we obtain 
much more explicite description of the arising process and we are
able to discuss its continuity properties. In particular, in the case
of Rindler wedge (Example 4.1.2) we show stochastic positivity of
the KMS structure arising in the context of Bisognano-Wichmann theorem
\cite{31}. These results can be used
to develop a rigorous constructive approach to non-linear quantum
field theories, by adopting the tools worked out in constructive
quantum field theory on a flat space-time . A detailed studies of
the perturbed Euclidean quantum fields
on stationary globally hyperbolic space-times will be presented in
our forthcoming publication. In this section we discuss also the models of
quantum statistical mechanics, including the model of
nonrelativistic Bose matter and infinite harmonic crystal, and describe
the corresponding stochastic processes.
\head II Weyl Algebra. Vacuum and Temperature Green Functions.
\endhead
\subhead
2.1 Weyl algebra
\endsubhead
Let $\cD$ be a real vector space with a locally convex topology $\tau$.
If $\sigma$ is a $\tau$- continuous symplectic form on $\cD$ (
i.e. $\sigma$ is a bilinear, antisymmetric and nondegenerate
mapping from $\cD \times \cD$ into $\bR$), then
$((\cD,\tau), \sigma)$ is called {\it a vector symplectic space}.
Let $W_{f}$ be the real function on $\cD$ defined by
$$
W_{f}(g)=\cases 1&\text{if}\: f=g\\
0&\text{if}\: f \neq g\endcases
$$
With the product
$$
W_{f}W_{g}=e^{-i\sigma(f,g)/2}\,W_{f+g}
$$
and involution
$$
W_{f}^{\ast}=W_{-f}
$$
the complex algebra generated by $W_{f}, f\in \cD$ becomes
a *-algebra $\fW_{0}(\cD,\sigma)$. We define $\WA$
(the {\it Weyl algebra over} 
$((\cD,\sigma)$) in the following way. Let 
$$
||\sum\limits_{k=1}^{N} z_{k}W_{f_{k}}||_{1}
=\sum\limits_{k=1}^{N}|z_{k}|
$$
$||\cdot||_{1}$ is a $\ast$-norm on $\fW_{0}(\cD,\sigma)$ and
let the completion $\conj{\fW(\cD,\sigma)}^{||\cdot||_{1}}$ 
is a $\ast$-Banach algebra with unit. Let ${\Cal F}$ be the set
of states on $\conj{\fW(\cD,\sigma)}^{||\cdot||_{1}}$, then
we define a C$^{\ast}$-algebra norm as follows \cite{32}:
$$
||W||=\sup\limits_{\rho\in {\Cal F}} \sqrt{\rho(W^{\ast}W)}
$$
$\WA$ is the completion of 
$\conj{\fW_{0}(\cD,\sigma)}^{||\cdot||_{1}}$ with respect to this
norm.
\par
A state $\omega$ on $\WA$ is called {\it regular} if it is 
$\tau$ - continuous i.e the function
$$
f \to \omega(W_{f})
$$
is $\tau$-continuous.
\remark{Remark}
Usually, the weaker form of continuity is assumed, namely
the map
$$
\bR \ni t \to \omega(W_{tf})
$$
should be continuous, for every $f\in \cD$.
\endremark
The set of all regular states on $\WA$ will be
denoted by $E(\fW)$. For a given $\omega \in E(\fW),
\GNS$ will be the corresponding GNS representation.
\par
Let $\{ T_{t} \}_{t\in \bR}$ be a one- parameter group of 
$\tau$- continuous symplectic mappings from $\cD$ onto $\cD$. In the
following, $(\cD,\sigma, \{ T_{t} \})$ will be called
{\it a linear dynamical system}. If we put
$$
\alpha_{t}(W_{f})=W_{T_{t}f}
$$
we obtain a one- parameter group of *- automorphisms of the Weyl algebra
$\WA$. For a given linear dynamical system  $\LDS$, let $E^{\alpha}(\fW)$
be the set of all regular $\alpha_{t}$- invariant states on $\WA$.
If $\omega \in E^{\alpha}(\fW)$, then on the GNS space ${\fH}_{\omega}$,
the evolution $\alpha_{t}$ is represented by a unitary group
$U_{\omega}(t)$ with a self-adjoint generator $H_{\omega}$.
\definition{Definition 2.1}
\roster
\item
$\omega \in E^{\alpha}(\fW)$ is {\it a ground state} iff $H_{\omega} \geq 0$.
\item
Let $0<\beta < \infty $ be given. A faithfull state 
$\omega\in E^{\alpha}(\fW)$
is  $\alpha_{t}$- {\it KMS state at the inverse temperature } $\beta$ 
iff in the corresponding 
GNS representation the unitary group $e^{itH_{\omega}}$ defines
a weakly continuous group of *- automorphisms of 
 $\pi_{\omega}(\WA)^{\prime \prime}$ such that
$$
\inprod{e^{-(\beta/2)H_{\omega}}\pi_{\omega}(W_{f})\Omega_{\omega}}
{e^{-(\beta/2)H_{\omega}}\pi_{\omega}(W_{g})\Omega_{\omega}}
=\inprod{\pi_{\omega}(W_{g})^{\ast}\Omega_{\omega}}
{\pi_{\omega}(W_{f})^{\ast}\Omega_{\omega}}
$$
for all $f,g \in \cD $. In this case $\Omega_{\omega}$ is cyclic
and separating for $\pi_{\omega}(\WA)^{\prime \prime}$ and the modular
operator associated with $\Omega_{\omega}$ coincides with 
$e^{-(\beta/2)H_{\omega}}$.
\endroster
\enddefinition
\subhead
2.2 Euclidean Green Functions.
\endsubhead
Let $\omega \in E^{\alpha}(\fW)$. We define the following {\it multi-time
Green functions}
$$
\align
\GF&=\omega(\alpha_{t_{1}}(W_{f_{1}})\cdots \alpha_{t_{n}}
(W_{f_{n}}))\\
&=\inprod{\Omega_{\omega}}{U_{\omega}(t_{1})\pi_{\omega}(W_{f_{1}})
U_{\omega}(-t_{1})\cdots U_{\omega}(t_{n})\pi_{\omega}(W_{f_{n}})
U_{\omega}(-t_{n}) \Omega_{\omega}}
\endalign
$$
If $\omega_{\infty} \in E^{\alpha}(\fW)$ is a ground state, then it can be 
shown that $\GF$ can be analytically continued to the tubular region
$$
{\frak T}^{\infty}_{n} = \{ (z_{1},\ldots,z_{n})\in {\bC}^{n}\, : \,
-\infty< \text{Im}\, z_{1}< \text{Im}\, z_{2}<,\ldots,< \text{Im}\,
z_{n}< \infty \}
$$
Let us define {\it the Euclidean region} ${\frak E}^{\infty}_{n}
\subset \conj{{\frak T}^{\infty}_{n}} $
$$
{\frak E}^{\infty}_{n} = \{ (s_{1},\ldots,s_{n})\in
\bR^{n}\, : \, s_{1}\leq,\ldots,\leq s_{n} \}
$$
The restriction of analytically continued Green functions to 
${\frak E}^{\infty}_{n}$ will be called {\it Euclidean Green functions}
corresponding to the ground state $\omega$ and denoted by 
$$
\EGF
$$
By 
linearity and continuity, we extend them to the Weyl algebra and obtain
$$
G^{E}(W_{1},\ldots,W_{n};s_{1},\ldots,s_{n}),\quad W_{1},\ldots,W_{n}
\in \WA
$$
Similarly, let $\omega \in E^{\alpha}(\fW)$ be the strongly $\beta$-KMS
state. By the theorem of Araki \cite{33}, the Green functions
$\GF$ can be analytically continued
to the  functions holomorphic in the tube
$$
{\frak T}^{\beta}_{n}=\{(z_{1},\ldots,z_{n})\in {\bC}^{n}\, : \,
-\beta/2<\text{Im}\,z_{1}<\cdots< \text{Im}\,z_{n}<\beta/2 \}
$$
and continuous on the boundary of ${\frak T}^{\beta}_{n}$. The restrictions
of holomorphic functions to the  {\it Euclidean region} defined by
$$
{\frak E}^{\beta}_{n} = \{ (s_{1},\ldots,s_{n})\, ; \,
-\beta/2 \leq s_{1} \leq\cdots\leq s_{n} \leq \beta/2 \}
$$
will be  called  {\it Euclidean Green functions}
corresponding to the KMS state $\omega$, and denoted by $\EGF$ as in the
ground state case. Again
by linearity and continuity, Euclidean Green functions defined by KMS
state
can be extended to the Weyl algebra $\WA$.
\par
The properties of Euclidean Green functions corresponding to the 
KMS state $\omega$ where studied in \cite{23}. Similarly we can obtain
the properties of ground state Euclidean Green functions. 
In the formulation of this properties, 
the following notation will be used: for $W_{1},\ldots,W_{k}\in
\WA$
$$
\Wk=(W_{1},\ldots, W_{k})\quad\text{and} \quad
\Wko=(W_{k}^{\ast},\ldots,W_{1}^{\ast})
$$
for $f_{1},\ldots,f_{k}\in {\cD},\,
\fk=(f_{1},\ldots,f_{k})
$. Similarly
$$
\sk=(s_{1},\ldots, s_{k})\quad\text{and}\quad
\sko=(-s_{k},\ldots, -s_{1})
$$
Moreover
$$
\align
G^{E}(\fk;\sk)&=\EGF\\
\EGW&=G^{E}(W_{1},\ldots,W_{k};s_{1},\ldots,s_{k})\\
\intertext{and}
\Wi&=(W_{1},\ldots,W_{i-1},W_{i}W_{i+1},W_{i+2},\ldots,W_{k})\\
\W{k-1}([i])&=(W_{1},\ldots,W_{i-1},W_{i+1},\ldots,W_{k})\\
\s{k-1}([i])&=(s_{1},\ldots,s_{i-1},s_{i+1},\ldots,s_{k})
\endalign
$$
For the ground state $\omega_{\infty}\in E^{\alpha}(\fW)$ the set of 
Euclidean Green functions
${\Bbb G}^{E}_{\infty}= \{ \EGW \}$ has 
the following properties:
\par
{\bf (EG1)}$_{\infty}$
\roster
\item
for each $\Wk$ the map  
$$
\sk \to \EGW
$$ 
is continuous,
\item
for a fixed $\sk\in \eo{k}$, the map 
$$
\Wk \to \EGW
$$ 
is multilinear and bounded i.e.
$$
|\EGW|\leq \prod\limits_{i=1}^{k}||W_{i}||
$$
Moreover, the map 
$$
\fk \to G^{E}(\fk;\sk)
$$ 
is $\tau$-continuous,
\item
for any $s\in [0,\infty)$
$$
G^{E}(\Wk;\sk +s)=\EGW,
$$
\item
for any $\Wk$ and $\sk\in \eo{k}$ such that 
$s_{i}=s_{i+1},\; 1\leq i \leq k-1$
$$
\EGW=G^{E}(\Wi,\s{k-1}([i+1]))
$$
\item
for any $\Wk$ such that $W_{i}=\bI:\; 1\leq i \leq k$
$$
\EGW=G^{E}(\W{k-1}([i]);\s{k-1}([i]))
$$
\item
$$
G^{E}({\Bbb I};0)=1
$$
\endroster
\par
{\bf (EG2)}$_{\infty}$ {\rm (OS-positivity)}
\par
For every terminating sequences
$$
(\W{0},\W{1},\ldots,\W{k},\ldots),\quad
(\s{0},\s{1},\ldots,\s{k},\ldots),\quad \s{k}\in {\frak E}^{\infty,+}_{k}
=\{ \s{k}\in\eo{k}\, : \, s_{1}\geq 0 \}
$$
\par 
we have
$$
\sum\limits_{k,l} G^{E}(\Wko,\W{l};\sko,\s{l})\geq 0
$$
\par
{\bf (EG3)}$_{\infty}$
\par
For every terminating sequences
$$
(\W{0},\W{1},\ldots,\W{k},\ldots),\quad
(\s{0},\s{1},\ldots,\s{k},\ldots),\quad \s{k}\in {\frak E}^{\infty,+}_{k}
=\{ \s{k}\in\eo{k}\, : \, s_{1}\geq 0 \}
$$
\par
and every $W \in \WA$ we have
$$
\sum\limits_{k,l}G^{E}(\Wko,W^{\ast},W,\W{l}\sko,0,0,\s{l})
\leq ||W||^{2}\sum\limits_{k,l}G^{E}(\Wko,\W{l};\sko,\s{l})
$$

Similarly, for $\beta$ - KMS state $\omega$ the set
${\Bbb 
G}^{E}=\{ \EGW \}$ of Euclidean Green functions has the following properties:
\par
{\bf (EG1)}$_{\beta}$
\roster
\item
for each $\Wk$ the map 
$$
\sk \to \EGW
$$
is continuous,
\item
for a fixed $\sk \in \eob{k}$ the map
$$
\Wk \to \EGW
$$
is multilinear, and bounded 
$$
|\EGW|\leq \prod\limits_{i=1}^{k}||W_{i}||
$$
Moreover, the map
$$
\fk \to G^{E}(\fk;\sk)
$$ 
is $\tau$-continuous,
\item
for any $\sk \in \eob{k}$ and $s\in [-\beta/2,\beta/2]$
such that $s_{k}+s \leq \beta/2$ the Euclidean Green functions
are locally shift invariant, i.e.
$$
G^{E}(\Wk;\sk +s)=\EGW
$$
\item
for any $\Wk$ and $\sk\in \eob{k}$ such that 
$s_{i}=s_{i+1},\; 1\leq i \leq k-1$
$$
\EGW=G^{E}(\Wi;\s{k-1}([i+1]))
$$

\item
for any $\Wk$ such that $W_{i}=1:\; 1\leq i \leq k$
$$
\EGW=G^{E}(\W{k-1}([i]);\s{k-1}([i]))
$$

\item
$$
G^{E}({\Bbb I};0)=1
$$
\endroster
\par
{\bf (EG2)}$_{\beta}$ {\rm (OS-positivity)}
\par
For every terminating sequences
$$
(\W{0},\W{1},\ldots,\W{k},\ldots),\quad
(\s{0},\s{1},\ldots,\s{k},\ldots),\quad \s{k}\in {\frak E}^{\beta,+}_{k}
=\{ \s{k}\in\eob{k}\, : \, s_{1}\geq 0 \}
$$
\par 
we have
$$
\sum\limits_{k,l} G^{E}(\Wko,\W{l};\sko,\s{l})\geq 0
$$
\par
{\bf (EG3)}$_{\beta}$
\par
For every terminating sequences
$$
(\W{0},\W{1},\ldots,\W{k},\ldots),\quad
(\s{0},\s{1},\ldots,\s{k},\ldots),\quad \s{k}\in {\frak E}^{\beta,+}_{k}
=\{ \s{k}\in\eob{k}\, : \, s_{1}\geq 0 \}
$$
\par
and every $W \in \WA$
we have
$$
\sum\limits_{k,l}G^{E}(\Wko,W^{\ast},W,\W{l};\sko,0,0,\s{l})
\leq ||W||^{2}\sum\limits_{k,l}G^{E}(\Wko,\W{l};\sko,\s{l})
$$
{\bf (EG4)}$_{\beta}$ {\rm (Weak form of KMS condition)}
\par
Let us define for $0\leq s_{1}\leq \cdots\leq s_{n}\leq\beta$
$$
\widehat{G}^{E}(W_{0},\ldots,W_{n};s_{1},\ldots,s_{n})
:=G^{E}(W_{0},\ldots,W_{n};-\beta/2,s_{1}-\beta/2,
\ldots,s_{n}-\beta/2)
$$
\par
Then for each $n$
$$
\widehat{G}^{E}(W_{0},\ldots,W_{n};s_{1},\ldots,s_{n})
$$
$$
=\widehat{G}^{E}(W_{n},W_{0},W_{1},\ldots,W_{n-1};
\beta-s_{n},\beta-s_{n}+s_{1},\ldots,\beta-s_{n}+s_{n-1})
$$
On the other hand, starting with Euclidean Green functions, and proceeding
similarly as in \cite{23} (see also \cite{18}), we can obtain
the following reconstruction theorems. Thus for ground state
case we have:
\proclaim{Theorem 2.1}
Let ${\Bbb G}^{E}_{\infty}$ be an abstract set of Euclidean Green functions 
on the Weyl algebra $\WA$ with properties {\rm (EG1)}$_{\infty}$--
{\rm (EG3)}$_{\infty}$. Then there exist:
\roster
\item
the Hilbert space ${\fH}_{\infty}$,
\item
the vector $\Omega_{\infty} \in {\fH}_{\infty}$,
\item
the $\ast$-representation $\pi_{\infty}$ of $\WA$ in 
${\Bbb B}({\fH}_{\infty})$,
\item
the weakly continuous one-parameter group of unitary operators
$U_{\infty}(t)=e^{itH_{\infty}}$ with $H_{\infty} \geq 0$
\endroster
such that
$$
G^{E}(\fk;\sk)=\langle\Omega_{\infty},e^{-s_{1}H_{\infty}}
\pi_{\infty}(W_{f_{1}})\cdots e^{-(s_{k}-s_{k-1})H_{\infty}}
\pi_{\infty}(W_{f_{k}})\Omega_{\infty}\rangle
$$
Moreover, the vector $\Omega_{\infty}$ is cyclic for the algebra
${\frak M}_{0}^{\infty}$ generated by 
$$e^{it_{1}H_{\infty}}\pi_{\infty}(W_{f_{1}})e^{-it_{1}H_{\infty}}
\cdots e^{it_{n}H_{\infty}}\pi_{\infty}(W_{f_{n}})e^{-it_{n}H_{\infty}};
\quad t_{1},\ldots,t_{n}\in \bR
$$
and the linear space spanned by vectors
$$
e^{-s_{1}H_{\infty}}\pi_{\infty}(W_{f_{1}})\cdots
e^{-(s_{k}-s_{k-1})H_{\infty}}\pi_{\infty}(W_{f_{k}})\Omega_{\infty}
$$
is dense in ${\fH}_{\infty}$. Additionally, $\alpha_{t}^{\infty}
(W)=e^{itH_{\infty}}We^{-itH_{\infty}}$ is $\sigma$-weakly continuous 
group of automorphisms of ${\frak M}^{\infty}=
({\frak M}_{0}^{\infty})^{\prime \prime}$.
\endproclaim
And similarly, for temperature case:
\proclaim{Theorem 2.2}
Let ${\Bbb G}^{E}_{\beta} $ be an abstract set of Euclidean Green functions
on the Weyl algebra $\WA$ with properties {\rm (EG1)}$_{\beta}$--
{\rm (EG4)}$_{\beta}$. Then there exist:
\roster
\item
the Hilbert space ${\fH}_{\beta}$,
\item
the vector $\Omega_{\beta}\in {\fH}_{\beta}$,
\item
the $\ast$-representation $\pi_{\beta}$ of $\WA$ in 
${\Bbb B}({\fH}_{\beta})$,
\item
the weakly continuous one-parameter group of unitary operators
$U_{\beta}(t)=e^{itH_{\beta}}$
\endroster
such that
$$
G^{E}(\fk;\sk)=\inprod{\Omega_{\beta}}
{e^{-s_{1}H_{\beta}}\pi_{\beta}(W_{f_{1}})\cdots 
e^{-(s_{k}-s_{k-1})H_{\beta}}\pi_{\beta}(W_{f_{k}})\Omega_{\beta}}
$$
Moreover, the vector $\Omega_{\beta}$ is cyclic for the algebra 
${\frak M}_{0}^{\beta}$ generated by operators of the form
$$
e^{it_{1}H_{\beta}}\pi_{\beta}(W_{f_{1}})e^{-it_{1}H_{\beta}}
\cdots e^{it_{n}H_{\beta}}\pi_{\beta}(W_{f_{n}})e^{-it_{n}H_{\beta}};
\quad t_{1},\ldots,t_{n} \in {\bR}
$$
and the state $\omega(M)=\inprod{\Omega_{\beta}}{M\Omega_{\beta}}$
is  $\beta$-KMS state on ${\frak M}^{\beta}
=({\frak M}_{0}^{\beta})^{\prime \prime}$
with respect to the unitary group $e^{itH_{\beta}}$.
\endproclaim
In the following, the W$^{\ast}-\beta$-KMS system constructed in
the Theorem 2.2 will be denoted as
$$
{\Bbb C}=(\fH_{\beta},\Omega_{\beta},\pi_{\beta},e^{itH_{\beta}},
{\frak M}^{\beta})
$$

\subhead 2.3 Stochastic positivity
\endsubhead
For a general KMS state $\omega$ on an abstract C$^{\ast}$-algebra $\frak 
A$, Klein and
Landau \cite{18} discussed the problem of construction of a stochastic
process corresponding to $\omega$. As they showed, such a process can
be constructed using some abelian sub-C$^{\ast}$-algebra $\frak B$ of 
$\frak A$ on which Euclidean Green functions corresponding to $\omega$
are positive in some special sense. In such case, the process has values
in the spectrum of the abelian algebra $\frak B$. To study the existence
of stochastic process in the case of Weyl algebra $\WA$, we consider
abelian subalgebras of $\WA$ defined in terms of so called 
{\it abelian splitting} of the symplectic space $(\cD,\sigma)$.
\definition{Definition 2.2}
A pair $({\cD}_{+},{\cD}_{-})$ of linear subspaces of the
symplectic space $({\cD},\sigma)$ is called an {\it abelian
splitting} if ${\cD}={\cD}_{+}+{\cD}_{-}$ and 
$\sigma({\cD}_{\pm},{\cD}_{\pm})=0$. For the abelian splitting
$({\cD}_{+},{\cD}_{-})$, let $\fW_{+}$ and $\fW_{-}$ be the
abelian subalgebras of the Weyl algebra $\WA$ generated by
$W_{f}$ with $f\in {\cD}_{+}$ and ${\cD}_{-}$ respectively.
\enddefinition
Now the desired positivity condition is the following:
\definition{Definition 2.3}
\roster
\item
The set of Euclidean Green functions ${\Bbb G}^{E}_{\beta}$
on the Weyl algebra $\WA$ with the given abelian splitting
is $\fW_{\pm}$-{\it stochastically positive} if for every
positive elements $W_{1},\ldots,W_{n} \in \fW_{\pm}$ and
every $(s_{1},\ldots,s_{n})\in \eob{n}$
$$
G^{E}(W_{1},\ldots,W_{n};s_{1},\ldots,s_{n})\geq 0
$$
\item
In the case of ground state system ${\Bbb G}^{E}_{\infty}$
the definition is analogous.
\endroster
\enddefinition
In the following discussion we will use the notions of thermal
and ground state processes with values in some Hausdorff topological
space ${\Bbb V}$.
\definition{Definition 2.4}
\roster
\item
Let ${\Bbb V}$ be a Hausdorff space. A stochastic process 
$\xi_{t}$ taking values in ${\Bbb V}$ will be called {\it thermal
process} (with the inverse temperature $\beta$) if:
\newline
TP1$_{\beta}$:
$\xi_{t}$ is stochastically continuous and faithful i.e.
for any $f\in \text{C}_{b}({\Bbb V});f>0$ we have ${\bE} f(\xi_{t})
\neq 0$,
\newline
TP2$_{\beta}$:
$\xi_{t}$ is periodic with period $\beta$ i.e. for all $t\in {\Bbb R}$,
$\xi_{t}=\xi_{t+\beta}$ in law (if $K_{\beta}$ is the circle with length
$\beta$ parametrized as $[-\beta/2,\beta/2]$ with end points glued, then 
$\xi_{t}$ can be indexed by $t\in K_{\beta}$),
\newline
TP3$_{\beta}$:
$\xi_{t}$ is OS-positive on $K_{\beta}$ i.e. for every 
$F \in \text{C}_{b}({\Bbb V}^{n})$ and $t_{1},\ldots,t_{n}\in [0,\beta/2]$
$$
{\bE}\conj{F}(\xi_{-t_{1}},\ldots,\xi_{-t_{n}})F(\xi_{t_{1}},\ldots,\xi_{t_{n}})
\geq 0
$$
\newline
TP4$_{\beta}$:
$\xi_{t}$ is shift invariant on $K_{\beta}$ i.e. $\xi_{t+\tau}
=\xi_{t}$ for all $t,\tau \in K_{\beta}$, and $t+\tau$ is defined modulo
$\beta$,
\newline
TP5$_{\beta}$:
$\xi_{t}$ is reflection invariant i.e. for every $F\in 
\text{C}_{b}({\Bbb V}^{n})$ and $t_{1},\ldots,t_{n}\in K_{\beta}$
$$
{\bE}F(\xi_{t_{1}},\ldots,\xi_{t_{n}})={\bE}F(\xi_{-t_{1}},\ldots,\xi_{-t_{n}})
$$
\item
Similarly, the process $\xi_{t}$ is called {\it ground state process}
if it is stochastically continuous and faithful, shift and reflection
invariant on $\bR$ and OS-positive on $\bR$.
\endroster
\enddefinition
Let $\xi_{t}$ be the thermal process with values in ${\cD}^{\ast}_{+}$,
where ${\cD}_{+}^{\ast}$ is an algebraic dual to ${\cD}_{+}$.
Define
$$
G^{\xi}(\fk;\s{k})={\bE}(\prod\limits_{l=1}^{k}e^{i<\xi_{s_{l}},f_{l}>})
$$
where $\fk\in {\cD}_{+}^{k}$ and $\s{k}\in K_{\beta}^{k}$.
It can be shown that ${\Bbb G}^{\xi}= \{ G^{\xi}(\fk;\s{k}) \}$
satisfy conditions (EG1)$_{\beta}$--(EG4)$_{\beta}$  for the abelian 
algebra ${\fW}_{+}$. Thus from Theorem 2.2, there exists the 
unique (up to the unitary equivalence) W$^{\ast}-\beta$-KMS system
$$
{\Bbb C}^{\xi}=({\fH}^{\xi},\Omega^{\xi},\pi^{\xi},e^{itH^{\xi}},
{\frak M}^{\xi})
$$
such that for every $\fk\in {\cD}_{+}^{k}$ and $\s{k}\in K_{\beta}^{k}$
$$
G^{\xi}(\fk;\s{k})=\inprod{\Omega^{\xi}}
{e^{-s_{1}H^{\xi}}\pi^{\xi}(W_{f_{1}})\cdots e^{-(s_{k}-s_{k-1})H^{\xi}}
\pi^{\xi}(W_{f_{k}})\Omega^{\xi}}
$$
Now we show the converse:
\proclaim{Theorem 2.3}
\roster
\item
Let $({\cD}_{+},{\cD}_{-})$ be the abelian splitting of the 
symplectic space $({\cD},\sigma)$. If ${\Bbb G}^{E}_{\beta}$ is 
the set of $\fW_{+}$-stochastically positive Euclidean Green functions
on $\WA$, there exists the unique (up to the stochastic equivalence)
thermal process $\xi^{\beta}_{t}$ with values in ${\cD}^{\ast}_{+}$
such that for every $f_{1},\ldots,f_{k} \in {\cD}_{+}$ and 
$\s{k}\in K_{\beta}^{k}$
$$
G^{E}(\fk;\s{k})={\bE}(\prod\limits_{l=1}^{k}e^{i<\xi_{s_{l}}^{\beta},f_{l}>})
\eqno (2.1)
$$
\item
Similarly, if ${\Bbb G}^{E}_{\infty}$ is the set of $\fW_{+}$-stochastically
positive ground state Euclidean Green functions, there exists the unique
(up to the stochastic equivalence) ground state process $\xi^{\infty}_{t}$
with values in ${\cD}^{\ast}_{+}$ such that for every $f_{1},\ldots,f_{k}
\in {\cD}_{+}$ and $-\infty < s_{1}\leq \cdots \leq s_{k}<\infty$
$$
G^{E}(\fk;\s{k})={\bE}(\prod\limits_{l=1}^{k}e^{i<\xi^{\infty}_{s_{l}},
f_{l}>})
\eqno (2.2)
$$
\endroster
\endproclaim
\demo{Proof}
Let ${\Bbb C}^{+}$ be the W$^{\ast}-\beta$-KMS system generated by
the set ${\Bbb G}^{+}$ of Euclidean Green functions $G^{E}(\fk;\s{k})$
restricted to $\fk \in {\cD}_{+}^{k}$. Then using the result of 
\cite{18}, we conclude that there exists the thermal process
$\xi^{KL}_{t}$ with values in the spectrum of von Neumann algebra
$(\pi_{+}(\fW_{+}))^{\prime \prime}$ (where $\pi_{+}$ denotes 
the representation defined by the set ${\Bbb G}^{+}$) and such that
$$
G^{E}(\fk;\s{k})=\int\limits_{Q_{\beta}}\prod\limits_{l=1}^{k}
\widehat{\pi_{+}(W_{f_{l}})}(q(s_{l}))d\mu^{\xi^{KL}}(q)
$$
where $Q_{\beta}$ is the path space of the process $\xi^{KL}$
and $\widehat{\pi_{+}(W_{f})}$ is the Gelfand transformation
of $\pi_{+}(W_{f})$.
\par
For fixed $\fk$ and $\s{k}$ define the map
$$
\pod{\alpha}\to \Gamma_{\fk,\s{k}}(\pod{\alpha})
$$
where $\pod{\alpha}=(\alpha_{1},\ldots,\alpha_{k})\in \bR^{k}$, by
$$
\Gamma_{\fk,\s{k}}(\pod{\alpha})=
G^{E}(\alpha_{1}f_{1},\ldots,\alpha_{k}f_{k};s_{1},\ldots,s_{k})
$$
$\Gamma$ is positive definite on $\bR^{k}$ since for every
$c_{l}\in\bC,\pod{\alpha}^{l}\in\bR^{k};l=1,\ldots,M$
$$\sum\limits_{l,m=1}^{M}c_{l}\conj{c_{m}}\Gamma_{\fk,\s{k}}
(\pod{\alpha}^{l}-\pod{\alpha}^{m})=
\int\limits_{Q_{\beta}}\vert \sum\limits_{l=1}^{M}\prod\limits
_{i=1}^{k}\widehat{\pi_{+}(W_{\alpha_{i}^{l}f_{i}})}(q(s_{i}))\vert^{2}
d\mu^{\xi^{KL}}(q)\geq 0
$$
By the Bochner theorem, there exists Borel probability measure 
$d\nu^{\fk,\s{k}}$ on $\bR^{k}$ such that
$$
\Gamma_{\fk,\s{k}}(\pod{\alpha})=
\int e^{i\inprod{\pod{\alpha}}{\pod{\gamma}}}d\nu^{\fk,\s{k}}
$$
The system of finite dimensional measures $d\nu^{\fk,\s{k}}$ forms
a compatible system of measures on the cylindrical sets of
${{\cD}_{+}^{\ast}}^{K_{\beta}}$ as it follows from (EG1)$_{\beta}$
(5).
Thus, by the theorem of Kolmogorov, there exists unique up to
the stochastic equivalence stochastic process $\xi^{\beta}_{t}$ with
values in ${\cD}_{+}^{\ast}$ such that (2.1) is satisfied. Similar
arguments can be used in the ground state case.
\enddemo
\remark{Remarks}
\roster
\item
It is worth to stress that the process corresponding to 
stochastically positive Euclidean Green functions on the Weyl
algebra $\WA$ can be realized as a process with
values not in an abstract spectrum
of abelian algebra, but in the dual space to $\cD_{+}$. In particular
sitauations, the space of values of the process $\xi^{\beta}_{t}$
can be localized in much smaller subspace of $\cD_{+}^{\ast}$ (
see i.e. Section IV below and \cite{24, 25}).
\item
In order to reconstruct all relevant informations about
KMS structure (or ground state structure) from the stochastic
process, it is important to take the abelian sub-algebra in such 
a way that the W$^{\ast}$-systems ${\Bbb C}$ and ${\Bbb C}^{\xi}$
(or analogous in the ground state case) are unitarily equivalent.
It is interesting that in the quasi-free case discussed in the
next section, abelian subalgebras defined by some natural
splitting always have this property.
\endroster
\endremark
\head III Quasi-free stochastically positive states.
\endhead
\subhead 3.1 Quasi-free states
\endsubhead
To discuss the properties of quasi-free stochastically positive systems,
assume that $\cD$ has a structure of complex pre-Hilbert space with
a scalar product $\inprod{\cdot}{\cdot}$ and $\sigma(f,g)= \text{Im}
\inprod{f}{g}$. Moreover, assume that $T_{t}$  is a group of
unitary operators with respect to $ \inprod{\cdot}{\cdot}$ which leaves
$\cD$ invariant and has 
a self - adjoint generator $\bold h$. 
Let ${\Bbb B}(f,g)$ be a positive, sesqulinear  form on ${\cD}$ which
is $T_{t}$-invariant and
such that
$$
|\sigma(f,g)|^{2}\leq{\Bbb B}(f,f)\,{\Bbb B}(g,g)
$$
Then $\omega$ defined by
$$
\omega(W_{f})=e^{-\frac{1}{4}{\Bbb B}(f,f)} \eqno (3.1)
$$
and extended  by linearity and continuity to the whole $\WA$ is a state
\cite{34}, so called {\it gauge invariant quasi-free state} on the Weyl
algebra $\WA$.

For quasi-free states, two point Green functions
$$
G^{(2)}(f,g:t)=\omega(W_{f}W_{T_{t}g})
$$
can be calculated, and have the form
$$
G^{(2)}(f,g;t)=e^{-\frac{1}{4}{\Bbb 
B}(f,f)-\frac{1}{4}{\Bbb B}(g,g)}e^{-\frac{1}{2}F(f,g;t)}\eqno (3.2)
$$
where
$$
F(f,g;t)=\text{Re}\,{\Bbb B}(f,T_{t}g)+i\sigma(f,T_{t}g)
$$
\subhead 3.2 KMS quasi-free states
\endsubhead
Now we consider a quasi-free state $\omega$ which is $\alpha_{t}$ -
KMS state at the inverse temperature $\beta$ for $\alpha_{t}$ defined by
$T_{t}=e^{it {\bold h}}$.
Then for every
$f,g \in \cD$ there exists a function $\varPhi(f,g;z)$ analytic
in the strip $0<\text{Im}\, z<\beta$ and continuous on the boundary,
such that
$$
\varPhi(f,g;t+i0)=F(f,g;t)
$$
and
$$
\varPhi(f,g;t+i\beta)=F(g,f;-t)
$$
Then
$$
G^{(2)}(f,g;t)=e^{-\frac{1}{4}({\Bbb B}(f,f)+{\Bbb B}(g,g))}
e^{-\frac{1}{2}\varPhi(f,g;t+i0)}
$$
and
$$
G^{(2)}(g,f;-t)=e^{-\frac{1}{4}({\Bbb B}(f,f)+{\Bbb B}(g,g))}e^{-\frac{1}{2}
\varPhi(f,g;t+i\beta)}
$$
To find the explicite form of ${\Bbb B}(f,g)$ let us first  consider the 
case when $\cD$ is a Hilbert space and ${\Bbb B}(f,g)$ is {\it bounded} on 
$\cD$, hence defined by some bounded operator $B$ such that $B\geq 1$.
\proclaim{Theorem 3.1}
Suppose that $\omega$ defined by (3.1) is an 
$\alpha_{t}$ - KMS state at the inverse temperature $\beta$. Then
\roster
\item
there exists $\varepsilon > 0$ such that $\bold h \geq \varepsilon $,
\item
$$
B=\frac{\bI+e^{-\beta {\bh}}}{\bI-e^{-\beta {\bh}}}
$$
\endroster
\endproclaim
\demo{Proof}
Since $\omega$ is $\alpha_{t}$ - KMS state, it is $\alpha_{t}$-invariant.
Thus $B$ commutes with $T_{t}=e^{it {\bold h}}$, hence $B$ commutes
with spectral projectors $E(\lambda)$ of $\bold h$.
First we show that $0$ is not an eigenvalue of $\bold h$. Suppose that
there exists $f_{0}\in \cD$ such that ${\bold h}f_{0}=0$.Then
for every $t\in \bR\, T_{t}f_{0}=f_{0}$. Because
$$
F(f,f_{0};t)=\text{Re}\,{\Bbb B}(f,T_{t}f_{0})+i \text{Im}\, 
\inprod{f}{T_{t}f_{0}}=F(f,f_{0};0)
$$
$\Phi(f,f_{0},z)$ 
is identically constant. Hence
$$
F(f,f_{0};0)=F(f_{0},f;0)
$$ 
On the other hand $F(f,f_{0};0)=\conj{F(f_{0},f;0)}$, what implies
$\sigma(f,f_{0})=0$. But $\sigma$ is nondegenerate, so $f_{0}=0$. 
Because $\Phi(f,f;z)$ is the analytic continuation of $F(f,f;t)$ and
$F(f,f;-t)$, so \cite{22}
$$
{\Cal F}({\Bbb T}_{F})(-p)=e^{-\beta p}{\Cal F}({\Bbb T}_{F})(p)
$$
where
${\Bbb T}_{F}$ is a distribution from ${\Cal S}^{\prime}$ given by
$$
<{\Bbb T}_{F},\varphi>=\int\limits_{- \infty}^{\infty} F(f,f,t)
\varphi(t)\,dt,\quad \varphi \in {\Cal S}
$$
and ${\Cal F}$ is the Fourier transform in ${\Cal S}^{\prime}$.
Because
$$
F(f,f;t)=\text{Re}\, \inprod{f}{Be^{it{\bold h}}f}+i
\text{Im}\, \inprod{f}{e^{it{\bold h}}f}
$$
$$
=\text{Re}\int\limits_{\bR\setminus \{ 0 \} } e^{it \lambda}\,
d\inprod{Bf}{E(\lambda)f}+i\text{Im}
\, \int\limits_{\bR\setminus \{ 0 \} } e^{it\lambda}\,
d\inprod{f}{E(\lambda)f}
$$
$$
= \frac{1}{2}\int\limits_{\bR\setminus \{ 0 \} }
e^{it \lambda}\,d \inprod{(B+\bI)f}{E(\lambda)f}+
\frac{1}{2}\int\limits_{\bR\setminus \{ 0 \} }
e^{-it\lambda}\, d\inprod{(B-\bI)f}{E(\lambda)f}
$$
for any $\varphi \in {\Cal S}$ we have
$$
<{\Cal F}({\Bbb T}_{F}),\varphi_{\text{inv}}>=
\frac{1}{2}\int\limits_{-\infty}^{\infty} dt
\int\limits_{\bR\setminus \{ 0 \} } \,d\inprod{(B+\bI)f}{E(\lambda)f}
e^{it\lambda} {\Cal F}(\varphi_{\text{inv}})(t)
$$
$$
+ \frac{1}{2}\int\limits_{-\infty}^{\infty} dt
\int\limits_{\bR\setminus \{ 0 \} }
\, d\inprod{(B-\bI)f}{E(\lambda)f}e^{-it\lambda} {\Cal F}
(\varphi_{\text{inv}})(t)
$$
$$
= \frac{\sqrt{2 \pi}}{2}
\int\limits_{\bR\setminus \{ 0 \} }
d\inprod{(B+\bI)f}{E(\lambda)f}\varphi(-\lambda)+
\frac{\sqrt{2 \pi}}{2}
\int\limits_{\bR\setminus \{ 0 \} }
d\inprod{(B-\bI)f}{E(\lambda)f} \varphi(\lambda) 
$$
On the other hand it is equal to
$$
<e^{-\beta \cdot} {\Cal F}({\Bbb T}_{F}),\varphi>
$$
$$
= \frac{\sqrt{ 2 \pi}}{2} \int\limits_{\bR\setminus \{ 0 \} }
d\inprod{(B+\bI)f}{E(\lambda)f}e^{-\beta \lambda}\varphi(\lambda)
+\frac{\sqrt{2 \pi}}{2}\int\limits_{\bR\setminus \{ 0 \} }
d\inprod{(B-\bI)f}{E(\lambda)f}e^{\beta\lambda}\varphi(-\lambda)
$$
Because $\text{Ker}\, {\bold h}=0$
$$
{\bold h}=\int\limits_{-\infty}^{0^{-}}\lambda dE(\lambda)+
\int\limits_{0^{+}}^{\infty} \lambda dE(\lambda)
$$
Assume that $E((-\infty,0))\neq 0$ and take $f\neq 0, \,
f \in \widetilde{\cD}=E((-\infty,0)){\cD}$. 
Suppose also that $\varphi \geq 
0$and $\text{supp}\, \varphi =[0,\infty)$. Then
$$
\int\limits_{-\infty}^{0} d\inprod{(B+\bI)f}{E(\lambda)f}\varphi(-\lambda)
=\int\limits_{-\infty}^{0} d\inprod{(B-\bI)f}{E(\lambda)f}e^{\beta\lambda}
\varphi(-\lambda)
$$
thus
$$
\inprod{f}{(B+\bI)\varphi({\bold h}^{-})f}=
\inprod{f}{(B-\bI)e^{-\beta {\bold h}^{-}}\varphi({\bold h}^{-})f}
$$
where
$$
{\bold h}^{-}=\int\limits_{-\infty}^{0} (-\lambda)dE(\lambda)
$$
Because $B\,:\, \widetilde{\cD}\to \widetilde{\cD}$ so
by polarization we obtain that
$$
(B+\bI)\varphi({\bold h}^{-})=(B-\bI)e^{-\beta {\bold h}^{-} }
\varphi({\bold h}^{-})
$$
But $\varphi$ is arbitrary, so
$$
B\vert_{\widetilde{\cD}}
=\frac{\bI+e^{-\beta {\bold h}^{-}}}{e^{-\beta 
{\bold h}^{-}}-\bI} <-\bI$$
Thus we get the contradiction. Hence ${\bold h} >0$ and using similar arguments
we show that
$$
B=\frac{\bI+e^{-\beta {\bold h}}}{\bI-e^{-\beta {\bold h}}}
$$
Suppose now that for every $n \in \bN$ the projector
$E((0,\frac{1}{n}))$ is nonzero. Take $f_{n}\in E((0,\frac{1}{n}))\cD$
such that $||f_{n}||=1$. Then
$$
||B^{1/2}f_{n}||^{2}=\int\limits_{0}^{1/n} a(\lambda)\,d\rho_{n}(\lambda)
$$
where
$a(\lambda)=(1+e^{-\beta \lambda})^{-1}(1-e^{-\beta \lambda})$
and
$d\rho_{n}(\lambda)=d\inprod{f_{n}}{E(\lambda)f_{n}}$. But for 
every $n\in \bN$ the measure $d\rho_{n}$ is normed to $1$ on 
$(0,\frac{1}{n})$,so
$$
||B^{1/2}f_{n}||^{2} \geq a(\frac{1}{n})\to \infty
$$
It contradictes the assumption that $B$ is bounded.

\enddemo
\subhead 3.3 Analytic continuation
\endsubhead
Starting with the functions $\varPhi(f,g;z)$ we can define two
point Euclidean Green function in terms of
$$
S(f,g;s)=\varPhi(f,g;is)\eqno (3.3)
$$
for $s\in [0,\beta]$. Then
$$
G^{E,2}(f,g;s)=e^{-\frac{1}{4}({\Bbb 
B}(f,f)+{\Bbb B}(g,g))}e^{-\frac{1}{2}S(f,g;s)}\eqno (3.4)
$$
\proclaim{Proposition 3.1}
$S(f,g;s)$ defined by (3.3) satisfy:
\roster
\item
For every $f,g \in \cD$ and $s\in [0,\beta]$
$$
S(f,g;s)=S(g,f;\beta-s)
$$
\item
For every $f\in \cD$ the mapping $s \to S(f,f;s)$
is OS-positive i.e. for every sequences $\{ s_{k} \}, s_{k}
\in [0,\beta/2];\{ c_{k} \}, c_{k} \in {\Bbb C}$
$$
\sum\limits_{k,l}\conj{c}_{k}c_{l}S(f,f;s_{k}+s_{l})\geq 0
$$
more generally, for every terminating sequences $\{ f_{k} \},
f_{k} \in {\cD}; \{ s_{k} \}, s_{k} \in [0,\beta/2]$ and 
$\{ c_{k} \}, c_{k} \in \bC$
$$
\sum\limits_{k,l}\conj{c}_{k}c_{l}S(f_{k},f_{l}; s_{k}+s_{l})
\geq 0
$$

\endroster
\endproclaim
\demo{Proof}
The function $S(f,g;s)$ can be represented by the following
integral formula \cite{22}
$$
S(f,g;s)= \int\limits_{-\infty}^{\infty} [{\frak P}(\rho,s)
F(f,g;\rho)+{\frak P}(\rho,\beta -s)F(g,f;-\rho)]d\rho
\eqno (3.5)
$$
with the kernel
$$
{\frak P}(\rho,s)=\frac{1}{2\beta}\sin \frac{\pi s}{\beta}(
\cosh \frac{\pi \rho}{\beta}-\cos \frac{\pi s}{\beta})^{-1}
$$
From this formula, we see that $S(f,g;s)=S(g,f;\beta-s)$.
On the other hand, the function $ t \to F(f,f;t)$ is positive definite
since
$$
F(f,g;t)=F_{+}(f,g;t)+F_{-}(f,g;t)
$$
where
$$
F_{+}(f,g;t)=\frac{1}{2}\inprod{f}{(B+\bI)e^{it {\bold h}}g}\,
\quad
F_{-}(f,g;t)=\frac{1}{2}\inprod{g}{(B-\bI)e^{-it {\bold h}}f}
$$
and the functions $F_{+}(f,f;t),\, F_{-}(f,f;t)$ are obviously positive
definite.
Thus $S(f,f;s)$ is OS-positive as a Laplace transform of positive 
measure. By polarization
$$
\sum\limits_{k,l}\conj{c}_{k}c_{l}F_{\pm}(f_{k},f_{l};t_{k}-t_{l}) \geq 0
$$
and
$$
F_{\pm}(f,g;t)=\int\limits_{- \infty}^{\infty} e^{it 
\lambda}\,d\mu^{\pm}_{f,g}(\lambda)
$$
for complex-valued measure $\mu^{\pm}_{f,g}$.  Thus
$$
\sum\limits_{k,l}\conj{c}_{k}c_{l}F(f_{k},f_{l};t_{k}-t_{l}) \geq 0
$$
and
$$
F(f,g;t)=\int\limits_{- \infty}^{\infty} e^{it\lambda}\,d\mu_{f,g}(\lambda)
$$
for $\mu_{f,g}=
\mu^{+}_{f,g}+\mu^{-}_{f,g}$ 

Hence the matrix of measures
$(\mu_{f_{k},f_{l}})$ is positive definite, and the matrix
$$
(S(f_{k},f_{l};s_{k}+s_{l}))=
(\int\limits_{-\infty}^{\infty} e^{-(s_{k}+s_{l})\lambda} \,
d\mu_{f_{k},f_{l}}(\lambda))
$$ 
is also positive-definite.
\enddemo
\proclaim{Corollary 3.1}
$$
G^{E,2}(f,g;s)
$$
satisfies:
\roster
\item
$$
G^{E,2}(f,g;s)=G^{E,2}(g,f;\beta-s)\quad;\quad s\in [0,\beta]
$$
\item
$$
\sum\limits_{k,l}\conj{c}_{k} c_{l} G^{E,2}(-f_{k},f_{l},s_{k}+s_{l})
\geq 0\quad ; \quad s_{k} \in [0,\beta/2]
$$
\endroster
\endproclaim
\remark{Remark}
Multi-time Euclidean Green functions corresponding to the
quasi-free KMS state can be obtained as follows. By a direct
computations we show that
$$
G(f_{1},\ldots,f_{n};t_{1},\ldots,t_{n})=
$$
$$
=\omega(W_{T_{t_{1}}f_{1}}\cdots W_{T_{t_{n}}f_{n}})=
\prod\limits_{l=1}^{n}e^{\frac{n-2}{4}{\Bbb B}(f_{l},f_{l})}
\prod\limits_{(j_{n},k_{n})} G^{(2)}(f_{j_{n}},f_{k_{n}};
t_{k_{n}}-t_{j_{n}})
$$
where the second product is taken over all pairs of different
indices $(j_{n},k_{n})$ such that $j_{n},k_{n}\in \{ 1,\ldots,n \},
t_{j_{n}} < t_{k_{n}}$. Analytic continuation of the right-hand side
of this expression gives after restriction to Euclidean points the
formula
$$
G^{E}(f_{1},\ldots,f_{n};s_{1},\ldots,s_{n})=
\prod\limits_{l=1}^{n}e^{\frac{n-2}{4}{\Bbb B}(f_{l},f_{l})}
\prod\limits_{(j_{n},k_{n})} G^{E,2}(f_{j_{n}},f_{k_{n}};
s_{k_{n}}-s_{j_{n}})
$$
which can be rewritten in the following way
$$
G^{E}(f_{1},\ldots,f_{n};s_{1},\ldots,s_{n})=
\prod\limits_{k=1}^{n}e^{- \frac{1}{4}{\Bbb B}(f_{k},f_{k})}
\prod\limits_{(j_{k},l_{k})}e^{- \frac{1}{2} S(f_{j_{k}},f_{l_{k}}
;s_{l_{k}}-s_{j_{k}})}
$$
It can be checked that $G^{E}(f_{1},\ldots,f_{n};s_{1},\ldots,s_{n})$
and satisfy all conditions of Theorem 2.1.
\endremark
\subhead 3.4 Stochastic positivity and thermal process
\endsubhead
Now we can pass to the question of stochastic positivity of
quasi-free Euclidean Green functions. This property can be 
naturally formulated in terms of the functions $S(f,g;s)$.
\par
Let $C$ be an abstract complex conjugation on $\cD$ i.e. $C$
is  antiunitary and $C^{2}=1$. The mapping $C$ naturally defines
the abelian splitting $({\cD}_{+},{\cD}_{-})$ of $\cD$ :
$$
{\cD}_{+}=\{ f\in {\cD}\,:\, Cf=f \}\quad,\quad
{\cD}_{-}=\{ f \in {\cD}\,:\, Cf=-f \}
$$
\proclaim{Lemma 3.1}
The functions $S(f,g;s)$ restricted to ${\cD}_{+}$ satisfy
$$
S(f,g;s)=S(g,f,;s),\quad s\in [0,\beta] \eqno (3.6)
$$
if $\bold h$ is $C$- real i.e. $C$ leaves the domain
$D(\bold h)$ invariant and 
$$
C{\bold h}f={\bold h}Cf
$$
for every $f \in D({\bold h})$
\endproclaim
\demo{Proof}
By analytic continuation, the property (3.6) is equivalent to
$$
F(f,g;t)=F(g,f;t)
$$
for $f,g \in {\cD}_{+}$ and $t \in \bR$. So we have to show that
$$
\text{Re}\,\inprod{f}{(T_{t}-T_{-t})Bg}=0 \eqno (3.7)
$$
and
$$
\text{Im}\inprod{f}{(T_{t}+T_{-t})g}=0 \eqno (3.8)
$$
for all $f,g \in {\cD}_{+}$.
Because $\bold h$ is $C$- real, $C$ commutes with
spectral projectors of $\bold h$ and
$$
CT_{t}=T_{-t}C
$$
thus
$$
T_{t}+T_{-t}\, : \, {\cD}_{+} \to {\cD}_{+}
$$
$$
T_{t}-T_{-t}\, : \, {\cD}_{+} \to {\cD}_{-}
$$
Hence (3.8) is satisfied. $B$ as a real function of $\bold h$
is $C$-real too
and so $B\,:\, \cD_{+}\to \cD_{+}$. But
$$
\text{Re}\inprod{f_{+}}{g_{-}}=0
$$
for all $f_{+}\in {\cD}_{+},\, g_{-}\in {\cD}_{-}$ so (3.7) follows.
\enddemo
Combining this results with that of Proposition 3.1 we get
$$
S(f,g;s)=S(f,g;\beta-s)\quad ; \quad s \in [0,\beta]
$$
This allows to extend the function $S(f,g;s)$ (for the fixed
$f,g\in {\cD}_{+}$) to the periodic function of $s$ with the period
$\beta$, defined for all $s\in \bR$. The extended functions will
also be denoted by $S(f,g;s)$.
\proclaim{Proposition 3.2}
For every fixed $f,g \in {\cD}_{+}$ there exists a finite,
real-valued measure $\nu_{f,g}$ on $[0,\infty)$ such that
$$
S(f,g;s)=\int\limits_{0}^{\infty}(e^{-sp}+e^{-(\beta-s)p})
\,d\nu_{f,g} \eqno (3.9)
$$
\endproclaim
\demo{Proof}
For a fixed $f\in {\cD}_{+}$, the function $S(f,f,s)$ is OS-positive,
thus by the Widder theorem there exists a finite measure
$\tilde{\nu}_{f,f}$ on $\bR$ such that
$$
S(f,f;s)=\int\limits_{-\infty}^{\infty} e^{-sp}\,d\tilde{\nu}_{f,f}
$$
for $s\in [0,\beta]$. Since $S(f,f;s)=S(f,f;\beta-s)$
$$
d\tilde{\nu}_{f,f}(-p)=e^{-\beta p}d\tilde{\nu}_{f,f}(p)
$$
Thus there exists a unique finite measure $\nu_{f,f}$ with support
on $[0,\infty)$ such that
$$
S(f,f;s)=\int\limits_{0}^{\infty}(e^{-sp}+e^{-(\beta-s)p})\,d\nu_{f,f}
$$
By polarization, this gives the integral representation of $S(f,g;s)$
in terms of real-valued measure $\nu_{f,g}$.
\enddemo
\proclaim{Proposition 3.3}
$S(f,g;s)$ is positive definite i.e.
$$
\sum\limits_{k,l=1}^{n}\conj{c}_{k}c_{l}S(f_{k},f_{l};s_{k}-s_{l})\geq 0
\eqno (3.10)
$$
for all terminating sequences $f_{1},\ldots,f_{n}\in {\cD}_{+}\,
;s_{1},\ldots,s_{n}\in \bR$ and $c_{1},\ldots,c_{n}\in \bC$.
\endproclaim
\demo{Proof}
Since 
$$
e^{-sp}+e^{-(\beta-s)p}=\sum\limits_{n \in \Bbb Z}
c_{n}(p) e^{i 2 \pi n s/\beta} \eqno (3.11)
$$
for all $s\in [0,\beta]$,
where
$$
c_{n}(p)=((p \beta)^{2}+(2 \pi n)^{2})^{-1}\,
(2 \beta p(1-e^{- \beta p})) \geq 0
$$
the right-hand side of (3.11) defines a periodic function 
for all $s\in \bR$ (\cite{35}). Combining this with (3.9) we get
$$
S(f,f;s)=\sum\limits_{n\in \Bbb Z}e^{i 2 \pi n s/\beta}
\int\limits_{0}^{\infty} c_{n}(p)\,d\nu_{f,f}(p)
$$
so for a fixed $f\in {\cD}_{+},\, S(f,f;s)$ is a real-valued
positive-definite function as a function of $s$. Thus,
it is a Fourier transform of a positive measure $m_{f,f}$ on
$\bR$. Using the polarization formula for $S(f,g;s)$ we obtain
that $S(f,g;s)$ is a Fourier transform
of real-valued measure $m_{f,g}$ with finite variation. Let
$$
m = \left( m_{f_{k},f_{l}} \right)
$$
be $n\times n$ matrix of such measures. We show that $m$ 
is positive-definite i.e. for every Borel subset $E\subset \bR$,
the matrix $\left( m_{f_{k},f_{l}}(E) \right)$
is positive-definite. Let $\varphi \in {\Cal S}$ be non-negative.
For $g_{k}=r_{k}f_{k},\, r_{k}\in \bR$ we have
$$
\sum\limits_{k,l}r_{k}r_{l}m_{f_{k},f_{l}}(\varphi)=
\sum\limits_{k,l}m_{g_{k},g_{l}}(\varphi)=
m_{(\sum\limits_{k}g_{k}),(\sum\limits_{k}g_{k})}(\varphi)\geq 0
$$
It implies that the matrix
$$
\left( S(f_{k},f_{l};s_{k}-s_{l})\right)=
\left( \int\limits_{- \infty}^{\infty} e^{i(s_{k}-s_{l})p}
dm_{f_{k},f_{l}}(p)\right) \eqno (3.12)
$$
is also positive-definite. To show this, notice that the
integral on the right-hand side of (3.12) can be approximated
by the sum
$$
\sum\limits_{j=0}^{N} e^{i(s_{k}-s_{l})p_{j}}
m_{f_{k},f_{l}}(E_{j})
$$
Because $\left( e^{i(s_{k}-s_{l})p_{j}}\right)$
and $\left(m_{f_{k},f_{l}}(E_{j})\right)$
are positive-definite matrices, from Schur lemma the matrix
$$
\left( e^{i(s_{k}-s_{l})p_{j}}m_{f_{k},f_{l}}(E_{j})\right)
$$
is positive-definite for each j, so the matrix (3.12) is positive-definite.
Thus $S(f_{k},f_{l};s_{k}-f_{l})$ is real-valued positive-definite matrix
in real vector space. It follows that it is also positive-definite
as a matrix in complex vector space.
\enddemo
\proclaim{Theorem 3.2}
$S(f,g;s)$ defines an operator-valued covariance function $R_{\beta}(s)$
of a periodic Gaussian OS-positive stochastic process indexed
by ${\cD}_{+}$. Thus $R_{\beta}(s)$ is an operator-valued
positive-definite function on ${\cD}_{+}$ which is periodic
and OS-positive. Moreover
$$
S(f,g;s)=\inprod{f}{R_{\beta}(s)g}
$$
\endproclaim
\demo{Proof}
From positive-definiteness of $S(f,g;s)$
we have
$$
|S(f,g;s)|^{2}\leq S(f,f;0)S(g,g;0)\leq ||B||^{2}\,||f||^{2}\,||g||^{2}
$$
Since $S(f,g;s)$ is bilinear and symmetric, there exists a
bounded and positive operator $R_{\beta}(s)$ on ${\cD}_{+}$ such that
$$
S(f,g;s)=\inprod{f}{R_{\beta}(s)g}
$$
Moreover, the function
$$
s\to R_{\beta}(s)
$$
is positive-definite, OS-positive and weakly periodic.

\enddemo
\proclaim{Corollary 3.2}
Let $\xi^{\beta}_{s}$ be the Gaussian process indexed by 
${\cD}_{+}$ defined by
$$
\bE(<\xi^{\beta}_{s},f>)=0
$$
and
$$
 \bE(<\xi^{\beta}_{s_{1}},f_{1}>
<\xi^{\beta}_{s_{2}},f_{2}>)=\frac{1}{2}\inprod{f_{1}}{R_{\beta}(s_{2}-s_{1})f_{2}}
$$
then, for $f_{1},f_{2}\in {\cD}_{+}$
$$
G^{E,2}(f_{1},f_{2};s_{2}-s_{1})= 
\bE(e^{i<\xi^{\beta}_{s_{1}},f_{1}>}\,e^{i<\xi^{\beta}_{s_{2}},f_{2}>})
$$
Similarly, for $f_{1},\ldots,f_{n}\in {\cD}_{+}$
$$
G^{E}(f_{1},\ldots,f_{n};s_{1},\ldots,s_{n})=
\bE(e^{i \sum\limits_{k=1}^{n}<\xi^{\beta}_{s_{k}},f_{k}>})
$$
\endproclaim
\proclaim{Proposition 3.4}
If $f \in {\cD}_{+}$ is such that  
$m(f)=\int\limits_{0}^{\infty}p\,d\nu_{f,f}(p)< \infty$,
then the coordinate process $\xi^{f}_{s}:=<\xi^{\beta}_{s},f>$ has a 
version (denoted by the same symbol) with H\"older continuous paths.
More precisely, for any $0<\gamma <1/2$ there exists an integrable
random variable $d(f,\gamma)$ such that
$$
|\xi^{f}_{s} - \xi^{f}_{s^{\prime}}| \leq d(f,\gamma)|s- s^{\prime}|^{\gamma}
$$
with probability one.
\endproclaim
\demo{Proof}
Since
$$
|S(f,f;h)-S(f,f;0)|=|\int\limits_{0}^{\infty}
((e^{-hp}-1)+(e^{-(\beta-h)p}-e^{-\beta p}))\,d\nu_{f,f}(p)|
$$
$$
\leq 2|h|\int\limits_{0}^{\infty} p\, d\nu_{f,f}(p)=2m(f)|h|
$$
it follows that
$$
|\bE \, e^{z|\xi^{f}_{s+h} - \xi^{f}_{s}|}|
\leq \bE\, e^{|z|(\xi^{f}_{s+h}-\xi^{f}_{s})} +
\bE\, e^{-|z|(\xi^{f}_{s+h} - \xi^{f}_{s})}
=2e^{\frac{|z|^{2}}{2}\bE(\xi^{f}_{s+h}-\xi^{f}_{s})^{2}}
$$
$$
\leq 2e^{|z||S(f,f;0)-S(f,f;h)|}\leq 2e^{2|z||h|m(f)}
$$
On the other hand, by the Cauchy integral formula
$$
|\xi^{f}_{s+h}-\xi^{f}_{s}|^{n}=\frac{n!}{2\pi i}
\int\limits_{C_{r}} \frac{e^{z|\xi^{f}_{s+h}-\xi^{f}_{s}|}}{z^{n+1}}
dz
$$
Thus
$$
\bE\,|\xi^{f}_{s+h}-\xi^{f}_{s}|^{n}\leq 2 n!
\frac{e^{2r^{2}|h|m(f)}}{r^{n}}
$$
Now taking $r=|h|^{-1/2}$ we obtain
$$
\bE |\xi^{f}_{s+h}-\xi^{f}_{s}|^{n}
\leq c_{n}|h|^{n/2}
$$
The assertion follows by the application of the Kolmogorov continuity test.
\enddemo

\definition{Definition 3.1}
A periodic stochastic process $\xi_{s}$ indexed by ${\cD}_{+}$ has
the {\it two-sided Markov property on the circle} $K_{\beta}$
iff for all $r,s \in K_{\beta}$
$$
E_{[s,r]}E_{[r,s]}=E_{ \{ r,s \} }E_{[r,s]}
$$
where for $I\subset K_{\beta},\, E_{I}$ denotes conditional expectation
with respect to the $\sigma$ -algebra generated by 
$\{ \xi_{s},\, s\in I \}$ and $ \{ r, s \} = \{ s, r \} $ is the set
consisting of the two elements $r,s$.
\enddefinition
\proclaim{Proposition 3.5}
$\xi^{\beta}_{s}$ has a version with the two-sided Markov property
on the circle $K_{\beta}$.
\endproclaim
\demo{Proof}
The covariance operator $R_{\beta}(s)$ has the form
$$
R_{\beta}(s)=\frac{e^{-s {\bold h}}+e^{-(\beta-s){\bold h}}}
{1-e^{-\beta {\bold h}}}
$$
Let us introduce the new scalar product in ${\cD}_{+}$ given by
$$
\inprod{f}{g}_{\beta}:=\inprod{f}{(1-e^{-\beta {\bold h}})^{-1}g}
$$
The norms $||\cdot||$ and $||\cdot||_{\beta}$ are obviously equivalent.
Let $\tilde{\xi}_{s}$ be the Gaussian process indexed by $({\cD}_{+},
\inprod{\cdot}{\cdot}_{\beta})$ with zero mean and covariance operator
$$
\tilde{R_{\beta}}(s)=e^{-s {\bold h}}+e^{-(\beta-s){\bold h}}
$$
$\tilde{\xi}_{s}$ is stochastically equivalent to $\xi^{\beta}_{s}$ and
by the result of \cite{35} it satisfies two-sided Markov property
on the circle $K_{\beta}$.
\enddemo
\subhead 3.5 KMS structure generated by thermal process
\endsubhead
Let $\xi^{\beta}_{s}$ be a Gaussian process constructed above and 
let $(Q,\Sigma,\mu)$ be its underlying probability space.
Since the process is stationary, $u(t)$ defined by
$$
u(t)(e^{i<\xi^{\beta}_{s_{1}},f_{1}>}\cdots 
e^{i<\xi^{\beta}_{s_{n}},f_{n}>})=e^{i<\xi^{\beta}_{s_{1}+t},f_{1}>}\cdots 
e^{i<\xi^{\beta}_{s_{n}+t},f_{n}>}$$
extends to a one parameter group of unitary operators on 
$L^{2}(Q,\Sigma,\mu)$. By periodicity, $u(\beta)=\bI$. 
Since the process is symmetric, $\Theta$ defined by
$$
\Theta (e^{i<\xi^{\beta}_{s_{1}},f_{1}>}\cdots e^{i<\xi^{\beta}_{s_{n}},
f_{n}>})=
e^{i<\xi^{\beta}_{-s_{1}},f_{1}>}\cdots e^{i<\xi^{\beta}_{-s_{n}},f_{n}>}
$$
extends to an unitary operator on $L^{2}(Q,\Sigma,\mu)$ such 
that $\Theta^{2}=\bI$. Finally, since the process is OS-positive$$
\inprod{\Theta F}{F}_{L^{2}} \geq 0
$$
for all $F \in L^{2}(Q,\Sigma_{[0,\beta/2]},\mu)$ where for $S\subset \bR$,
$\Sigma_{S}$ denotes the $\sigma$-algebra generated by $\{ \xi^{\beta}_{s}
\,:\, s\in S \}$.
\proclaim{Theorem 3.3}
Let $\xi^{\beta}_{s}$ be a Gaussian, periodic (with period $\beta$),
OS-positive stochastic process
indexed by ${\cD}_{+}$. Then there exist a Hilbert space
$\fH_{\xi}$ with a unit vector $\Omega_{\xi}$, a weakly continuous
one parameter group of unitary operators $U_{\xi}(t)=e^{it H_{\xi}}$
and a von Neumann algebra $\fM_{\xi}$ of operators acting
on $\fH_{\xi}$ such that $\Omega_{\xi}$ is cyclic and separating
for $\fM_{\xi}$ and $\alpha^{\xi}_{t}(M)=e^{it H_{\xi}}M e^{-it H_{\xi}}$ 
is the modular automorphisms group associated with $\Omega_{\xi}$.

\endproclaim
\demo{Proof}
On the space $L^{2}(Q,\Sigma_{[0,\beta/2]},\mu)$ define a sesquilinear
form by
$$
\inprod{F}{G}=\inprod{\Theta F}{G}_{L^{2}}
$$
By OS-positivity, it is positive semi-definite. Let
$$
\cN= \{ F\in L^{2}(Q,\Sigma_{[0,\beta/2]},\mu)\,:\,
\inprod{F}{F}=0 \}
$$
Then 
$$
\fD=L^{2}(Q,\Sigma_{[0,\beta/2]},\mu)/\cN
$$
is a pre-Hilbert space with respect to the inner product
$$
\inprod{[F]}{[G]}=\inprod{F}{G}
$$
where $[F]$ denotes the class containing $F$. $\fH_{\xi}$ is defined
as a Hilbert space completion of $\fD$ and $\Omega_{\xi}=[1]$. 
Let $\fD_{t}$ be the linear space
generated by vectors $[F]$ for $F\in L^{2}(Q,\Sigma_{[0,\beta/2- 
	t]},\mu),\,t \in [0, \beta/2]$. For every $t \in [0,\beta/2]$ we can 
define the linear operator $p(t)$ with domain $\fD_{t}$ by
$$
p(t)[F]=[u(t)F]
$$
and we can show that $(p(t),\fD_{t})$ form a symmetric local semigroup
(\cite{36}). Hence there exists a unique self-adjoint operator
$H_{\xi}$ on $\fH_{\xi}$ such that $p(t)=e^{-t H_{\xi}}$. $U_{\xi}(t)$ is 
defined by $U_{\xi}(t)=e^{it H_{\xi}}$. Let now $F_{0}\in L^{\infty}
(Q,\Sigma_{0},\mu)$. Then
$$
\pi_{0}(F_{0})[F]=[F_{0}F]
$$
defines a bounded operator on $\fH_{\xi}$ and 
$$
\fM_{0}= \{ \pi_{0}(F_{0})\,:\, F_{0}\in L^{\infty}(Q,\Sigma_{0},\mu) \}
$$ 
is a von Neumann algebra of operators on $\fH_{\xi}$.
Let $\fM_{\xi}$ be the von Neumann algebra generated by elements
$$
e^{it_{1} H_{\xi}}B_{1}e^{-it_{1} H_{\xi}}\cdots e^{it_{n} H_{\xi}}
B_{n} e^{-it_{n} H_{\xi}}
$$ 
with $t_{j}\in \bR, B_{j} \in \fM_{0}$. Then $\Omega_{\xi}$ is cyclic and
separating for $\fM_{\xi}$ (\cite{23}). Using the properties of the
process $\xi^{\beta}_{s}$ we can now define the modular conjugation
and modular group corresponding to $\Omega_{\xi}$. Let
$$
\Theta_{\beta/4}=u(\beta/4)\Theta u(-\beta/4)
$$
and
$$
J_{\xi}[F]=[\conj{\Theta_{\beta/4}F}]
$$
Then
$$
\inprod{J_{\xi}[F]}{J_{\xi}[G]}=\inprod{\Theta \conj{\Theta_{\beta/4}F}}
{\conj{\Theta_{\beta/4}G}}_{L^{2}}=
\inprod{G}{\Theta F}_{L^{2}}=\inprod{[G]}{[F]}
$$
since $\Theta_{\beta/4}$ commutes with $\Theta$ by periodicity. Hence 
$J_{\xi}$ can be extended to an antiunitary operator on $\fH_{\xi}$ such that
$J_{\xi}^{2}=\bI$. Computing the action of $J_{\xi}$ on 
$\fM_{\xi}\Omega_{\xi},\, M\in \fM_{\xi}$ we can show that
$$
J_{\xi}M^{\ast}\Omega_{\xi}= e^{(-\beta/2)H_{\xi}}M\Omega_{\xi}
$$
Now since $\fM_{\xi}\Omega_{\xi}$ is a core for $e^{(-\beta/2)H_{\xi}}$
(\cite{18}), $J_{\xi}$ defined above is the modular conjugation operator
and the corresponding modular operator $\Delta_{\xi}$ can be identified with 
$e^{(-\beta/2)H_{\xi}}$. For more details see \cite{18}.
\enddemo
Let $(\fH_{\omega},\pi_{\omega},\Omega_{\omega})$ be the GNS representation
defined by quasi-free KMS state $\omega$. Then $\Omega_{\omega}$ is 
cyclic and separating for $\pi_{\omega}(\WA)^{\prime \prime}$. Let 
$J_{\omega}$ and $\Delta_{\omega}$ be the corresponding modular conjugation 
and modular operator. We are going to show that in the case of quasi-free 
state the modular structure constructed from the process $\xi^{\beta}_{s}$
is unitarily equivalent to the canonical modular structure defined
by KMS state $\omega$. Thus all relevant informations about KMS structure
are contained in the (commutative) stochastic process $\xi^{\beta}_{s}$.
To obtain this result we need the following property of the dynamics
$e^{it\bh}$.

\proclaim{Theorem 3.4}
Let  $\bh$ be $C$-real on $\cD$ and let $E(\{ 0 \})=0$, where $dE(\lambda)$
is the spectral measure of $\bh$. Then
the family 
$$
\{ \sum\limits_{k=1}^{n} e^{it_{k}\bh}f_{k}\,:\, n\in \bN,t_{k}\in
\bR, f_{k}\in \cD_{+} \}
$$
is dense in $\cD$.

\endproclaim
\demo{Proof}
Because $C$ commutes with spectral projectors of $\bh$ and $0$ is not
an eigenvalue of $\bh$, so
$$
\bh= \bh_{1}\oplus \bh_{2},\quad \cD=\cD_{1}\oplus \cD_{2}\quad
\text{and}\quad \cD_{+}=\cD_{+,1}\oplus \cD_{+,2}
$$
where
$$
\cD_{1}=E((0,\infty))\cD,\quad \cD_{2}=E((-\infty,0))\cD,\quad
\cD_{+,1}=\cD_{+}\cap \cD_{1},\quad \cD_{+,2}=\cD_{+}\cap \cD_{2}
$$
Hence it is enough to consider a positive operator $\bh$ such that
$\text{Ker}\, \bh=\{ 0 \}$. Because $\cD$ is separable, $\bh$ can be
written as
$$
\bh=\oplus_{l=1}^{\infty} \bh_{l},\quad
\cD=\bigoplus_{l=1}^{\infty} \cD_{l}
$$
where each $\bh_{l}$ is a positive operator with a simple spectrum. 
Moreover, each generating vector $g_{l}$ for a Hilbert space $\cD_{l}$
can be taken from $\cD_{+}$ and normalized. Indeed, let us take arbitrary
$g_{1}\in \cD_{+}$ with $||g_{1}||=1$ and define
$$
\cD_{1}=\conj{\text{Lin}_{\bC}\, \{ E(\sigma)g_{1} \}},\quad
\sigma \in \cB(0,\infty)
$$
Because $C$ commutes with all $E(\sigma)$, so $C\,:\, \cD_{1}\to \cD_{1}$
and hence $C \,:\, \cD_{1}^{\bot}\to \cD_{1}^{\bot}$. In the second step
we take $g_{2}\in \cD_{1}^{\bot}\cap \cD_{+},\, ||g_{2}||=1$ and define
$$
\cD_{2}=\conj{\text{Lin}_{\bC}\, \{ E(\sigma) g_{2} \}}
$$
It is clear that $\cD_{1}\bot \cD_{2}$. The total decomposition
follows from Kuratowski-Zorn lemma. Because every $\cD_{l}$ reduces
$\bh$, $\bh=\oplus_{l=1}^{\infty} \bh_{l}$ where $\bh_{l}\,:\,
D(\bh)\cap \cD_{l} \to \cD_{l}$. Moreover, each $\bh_{l}$ is
$C\vert_{\cD_{l}}$-real.
\par
Let us assume that we have proved that for every $n\in \bN$ the set
$$
\{ \sum\limits_{k=1}^{m} e^{it_{k}\bh_{k}}f_{n}^{k}\,:\, m\in \bN,
t_{k}\in \bR, f_{n}^{k} \in \cD_{+,n} \}
$$
is dense in $\cD_{n}$. Then for any $\varepsilon >0$ and for any $f\in 
\cD$ we can find $n_{0}\in \bN$ such that
$$
||\bigoplus_{n=1}^{n_{0}} f_{n} -f||<\varepsilon/3
$$
and for these   $f_{n}\in \cD_{n}$ there are $f_{n}^{k} \in \cD_{+,n}$
such that
$$
||f_{n}-\sum\limits_{k=1}^{m_{n}} e^{it_{k}(n)\bh_{n}}f_{n}^{k}||\leq
\varepsilon/2^{n}
$$
It follows that
$$
||f-\bigoplus_{n=1}^{n_{0}}\sum\limits_{k=1}^{m_{n}} 
e^{it_{k}(n)\bh_{n}}f_{n}^{k}||\leq 
||f-\bigoplus_{n=1}^{n_{0}}f_{n}||+||\bigoplus_{n=1}^{n_{0}} f_{n}-
\bigoplus_{n=1}^{n_{0}}\sum\limits_{k=1}^{m_{n}} e^{it_{k}(n)\bh_{n}} 
f_{n}^{k}||$$
$$
\leq \varepsilon/3 + \left[\sum\limits_{n=1}^{n_{0}} || f_{n} 
-\sum\limits_{k=1}^{m_{n}}e^{it_{k}(n)\bh_{n}} f_{n}^{k}||^{2}\right]^{1/2}
\leq \varepsilon/3 
+\left[\sum\limits_{n=1}^{n_{0}}(\varepsilon^{2}/4^{n})\right]
^{1/2}<\varepsilon
$$
Because every $f_{n}^{k}$ is an element of $\cD_{+}$ and we can write
$$
e^{it_{k}(n)\bh_{n}}f_{n}^{k}=e^{it_{k}(n)\bh} f_{n}^{k}
$$
we obtain
$$
||f-\sum\limits_{n=1}^{n_{0}}\sum\limits_{k=1}^{m_{n}}e^{it_{k}(n)\bh}
f_{n}^{k}||<\varepsilon
$$
Hence to finish the proof we need only the following Lemma.
\enddemo
\proclaim{Lemma 3.2}
Suppose that $\bh$ is $C$-real and positive on $\cD$. If $\text{Ker}\,\bh
=\{ 0 \}$ and $\bh$ is simple with a generating vector $g\in \cD_{+},\,
||g||=1$, then the family
$$
\{ \sum\limits_{k=1}^{n} e^{it_{k}\bh} f_{k}\,:\, n\in \bN, t_{k}\in \bR,
f_{k} \in \cD_{+} \}
$$ 
is dense in ${\cD}$.
\endproclaim
\demo{Proof}
By the spectral theorem, $\bh$ is unitarily isomorphic to the multiplication
operator $\widehat{x}$ in $L^{2}(\bR_{+},\cB,\rho)$, where
$\rho(\sigma)=\inprod{g}{E(\sigma)g}$. It means that the unitary operator
$U$ given by
$$
Ua=\int\limits_{0}^{\infty} a(\lambda)\,dE(\lambda)g
$$
maps $D(\widehat{x})$ onto $D(\bh)$ and $U^{\ast}\bh U=\widehat{x}$.
Because $C$ commutes with $E(\sigma)$ and $Cg=g$, we also have
that $U(L^{2}_{r})=\cD_{+}$ where
$$
L^{2}_{r} = \{ a \in L^{2}(\bR_{+},\cB,\rho)\,:\, \conj{a}=a \}
$$
Thus for $a_{k}\in L^{2}_{r}$
$$
U(\sum\limits_{k=1}^{n} e^{it_{k} \widehat{x}} a_{k})=
\sum\limits_{k=1}^{n} e^{it_{k} U\widehat{x}U^{\ast}}(Ua_{k})=
\sum\limits_{k=1}^{n} e^{it_{k}\bh}f_{k}
$$
where $f_{k}=Ua_{k}$. So it is enough to proof the statement for
$\widehat{x}$.
\par
Let $K$ be an arbitrary compact set $K\subset (0, \infty)$. Let
us consider the family of continuous, real valued functions on
$K$ given by
$$
\{ a\,:\, a=\sum\limits_{k=1}^{n} \sin (t_{k} \widehat{x}) a_{k},
n\in \bN, a_{k} \in C_{r}(K) \}
$$
It is clear that this family is an algebra which separates points in
$K$ and does not vanish identically on some point in $K$. By 
Stone-Weierstrass theorem this family is dense in $C_{r}(K)$.
Now let $f\in C(K)$. Then $f=f_{1}+i f_{2}, f_{1},f_{2} \in C_{r}(K)$.
At first we find a finite sum $\sum\limits_{k=1}^{n} \sin (t_{k} \widehat{x} )
a_{k}, a_{k}\in C_{r}(K)$ such that
$$
||f_{2}-\sum\limits_{k=1}^{n} \sin (t_{k}\widehat{x}) a_{k} ||_{\text{sup}}
< \varepsilon
$$
 Next put $t_{0}=0$, and $a_{0}=f_{1}-\sum\limits_{k=1}^{n} \cos (t_{k}\widehat{x})
a_{k}$. Then
$$
||f-\sum\limits_{k=0}^{n} e^{it_{k}\widehat{x}}a_{k}||_{\text{sup}}=
||f_{2}-\sum\limits_{k=1}^{n} \sin (t_{k}\widehat{x}) a_{k}||_{\text{sup}}
<\varepsilon
$$
Because $\rho$ is a probability measure on $\bR_{+}$, we have also that
$$
||f-\sum\limits_{k=0}^{n} e^{it_{k}\widehat{x}}a_{k}||_{L^{2}} <\varepsilon
$$
But $\bR_{+}$ is locally compact and $\rho$ is a Borel measure, so for every
$\varphi\in L^{2}(\bR_{+},\cB,\rho)$ there is a compact set $K\subset 
\bR_{+}$ and a function $f\in C(K)$ such that $||\varphi -f||_{L^{2}} <
\varepsilon$. Thus the proof of Lemma is finished.
\enddemo
\proclaim{Theorem 3.5}
For quasi-free KMS state $\omega$ defined on the Weyl algebra
$\WA$ the canonical modular structure $(\pi_{\omega}(\WA)
^{\prime \prime},\Delta_{\omega},J_{\omega})$ is unitarily
equivalent to the modular structure $(\fM_{\xi},\Delta_{\xi},
J_{\xi})$ constructed from
the stochastic process $\xi^{\beta}_{s}$.
\endproclaim
\demo{Proof}
Let $w(f)$ denote the following elements of $L^{\infty}(Q,\Sigma_{0},\mu)$
$$
w(f)=e^{i<\xi^{\beta}_{0},f>}
$$
On a dense set of vectors in $\fH_{\xi}$ generated by elements
$$
e^{it_{1}H_{\xi}}\pi_{0}(w(f_{1}))e^{-it_{1}H_{\xi}}
\cdots e^{it_{n}H_{\xi}}\pi_{0}(w(f_{n}))e^{-it_{n}H_{\xi}}
\Omega_{\xi},\quad t_{1},\ldots,t_{n}\in \bR,f_{1},\ldots,f_{n}
\in \cD_{+}
$$
we define a map $V$ by
$$
V\left( e^{it_{1}H_{\xi}}\pi_{0}(w(f_{1}))e^{-it_{1}H_{\xi}}
\cdots 
e^{it_{n}H_{\xi}}\pi_{0}(w(f_{n}))e^{-it_{n}H_{\xi}}\Omega_{\xi}\right)
$$
$$
=e^{it_{1}H_{\omega}}\pi_{\omega}(W_{f_{1}})e^{-it_{1}H_{\omega}}
\cdots e^{it_{n}H_{\omega}}\pi_{\omega}(W_{f_{n}})e^{-it_{n}H_{\omega}}
\Omega_{\omega}
$$
Since the functions
$$
\phi_{\xi}(t_{1},\ldots,t_{n})=\inprod{\Omega_{\xi}}
{e^{it_{1}H_{\xi}}\pi_{0}(w(f_{1}))e^{-it_{1}H_{\xi}}
\cdots 
e^{it_{n}H_{\xi}}\pi_{0}(w(f_{n}))e^{-it_{n}H_{\xi}}\Omega_{\xi}}
$$
and
$$
\phi_{\omega}(t_{1},\ldots,t_{n})=\inprod{\Omega_{\omega}}
{e^{it_{1}H_{\omega}}\pi_{\omega}(W_{f_{1}})e^{-it_{1}H_{\omega}}
\cdots e^{it_{n}H_{\omega}}\pi_{\omega}(W_{f_{n}})e^{-it_{n}H_{\omega}}
\Omega_{\omega}}
$$
can be analytically continued to ${\frak T}^{\beta}_{n}$ and coincide
in Euclidean points, they are equal for all $t_{1},\ldots,t_{n} \in
\bR$. Thus $V$ is an isometry with dense domain in $\fH_{\xi}$. But
$$
e^{it_{1}H_{\omega}}\pi_{\omega}(W_{f_{1}})e^{-it_{1}H_{\omega}}
\cdots e^{it_{n}H_{\omega}}\pi_{\omega}(W_{f_{n}})e^{-it_{n}H_{\omega}}
$$
$$
= \pi_{\omega}(W_{\sum\limits_{k=1}^{n} e^{it_{k}\bh}f_{k}})
\prod\limits_{1\leq k < l \leq n}   
e^{-\frac{i}{2}\sigma(f_{k},e^{i(t_{l}-t_{k})\bh}f_{l})}
$$
hence the range of $V$ is dense in $\fH_{\omega}$ by Theorem 3.4
and continuity of $f\to \omega(W_{f})$ with respect to the Hilbert space
norm on $\cD$. 
Moreover  
$V\Omega_{\xi}=\Omega_{\omega}$. By direct computaion we show that
$$
V\pi_{0}(w(f))V^{\ast}=\pi_{\omega}(W_{f}),\quad f\in \cD_{+};
\quad Ve^{itH_{\xi}}V^{\ast}=e^{itH_{\omega}}
$$
and
$$
V\left(\prod\limits_{k=1}^{m} e^{it_{k}H_{\xi}}\pi_{0}(w(f_{k}))
e^{-it_{k}H_{\xi}}\right)V^{\ast}=
\prod\limits_{k=1}^{m} e^{it_{k}H_{\omega}}\pi_{\omega}(W_{f_{k}})
e^{-it_{k}H_{\omega}}
$$ 
Thus
$$
V\fM_{\xi}V^{\ast}=\pi_{\omega}(\WA)^{\prime \prime}
$$
and 
$$
VJ_{\xi}V^{\ast}=J_{\omega},\quad V\Delta_{\xi}V^{\ast}=\Delta_{\omega}
$$

\enddemo
\subhead 3.6 General case
\endsubhead
If the operator $\bh$ is only bounded from below, that is
$$
{\bh}\geq \mu {\bI},\quad \mu <0
$$
then we can still obtain a quasi-free KMS state on the whole
algebra $\WA$ if we replace $\bh$ by $\tilde{\bh}=
{\bh}-(\mu-1){\bI}$ for example. Then the operator $B$ will be equal to
$$
B=\frac{\bI +z e^{-\beta \bh}}{\bI- z e^{-\beta \bh}}
$$
where $z=e^{\beta (\mu-1)}$ and $B$ will be bounded. But at the same time
the dynamics $\alpha_{t}$ have to be repleced by $\tilde{\alpha}_{t}$
given by
$$
\tilde{\alpha}_{t}(W_{f})=W_{e^{it \tilde{\bh}}f}=W_{z^{-it/\beta} T_{t}f}
$$
In some cases it is necessary to consider the operator $\bh$ itself. Then
we have to restrict $\WA$ to a Weyl algebra over a suitable subspace of 
the Hilbert space $\cD$.
\proclaim{Theorem 3.6}
Suppose that $\bh \geq 0$ and $\text{Ker}\,\bh={0}$. Let 
$
B=(\bI+e^{-\beta \bh})^{-1}(\bI+e^{-\beta \bh})$ and 
$$
{\Bbb B}(f,g)=
\inprod{B^{1/2}f}{B^{1/2}g} \quad \text{with}\quad D({\Bbb B})=D(B^{1/2})
$$ 
Let us define
$$
\omega(W_{f})=e^{-\frac{1}{4} {\Bbb B}(f,f)} \quad \text{for}
\quad f\in D({\Bbb B})
$$
and extend it onto the Weyl algebra $\fW(D({\Bbb B}),\sigma)$.Then $\omega$ is an
$\alpha_{t}$-KMS state at the inverse temperature $\beta$ for the dynamics 
$\alpha_{t}$ given by $T_{t}=e^{it\bh}$.
\endproclaim
\demo{Proof}
First let us check that $\alpha_{t}$ maps $\fW(D({\Bbb B}),\sigma)$ into itself.
But since $T_{t}$ commutes with spectral projectors of $\bh$
$$
T_{t}\, : \, D(B^{1/2})\to D(B^{1/2})
$$
hence $\alpha_{t}$ leaves $\fW(D({\Bbb B}),\sigma)$ invariant. Now consider
two point Green function $G^{(2)}(f,g;t)$ given by (3.2) for $f,g\in
D({\Bbb B})$. It is clear that this function is continuous and bounded. Hence to
show the existence of analytic function interpolating between 
$G^{(2)}(f,g;t)$ and $G^{(2)}(g,f;-t)$ in the strip $\bR\times i[0,\beta]$
it is enough to check that
$$
\cF(\bT_{F(f,g;\cdot)})=e^{-\beta p}\cF(\bT_{F(g,f;\cdot)})
$$
For $f=g$ it can be done by direct calculations similarly as in the 
proof of Theorem 3.1. The general result follows from the 
polarization formula. It is clear that such analytic function exists
for any 
$$
a_{n}=\sum\limits_{k=1}^{n}c_{k}W_{f_{k}},\quad
b_{n}=\sum\limits_{k=1}^{n}d_{k}W_{g_{k}}
$$
in $\fW(D({\Bbb B}),\sigma)$. Now for arbitrary $a,b \in
\fW(D({\Bbb B}),\sigma)$
$$
G^{(2)}(a,b;t)=\lim\limits_{n\to \infty} G^{(2)}(a_{n},b_{n};t)
$$
and this function is continuous and bounded, so
$$
\bT_{G^{(2)}(a,b;\cdot)}=\lim\limits_{n\to \infty} \bT_{G^{(2)}(a_{n},b_{n};
\cdot)}
$$
in ${\Cal S}^{\prime}$. But since ${\cF}$ is a homeomorphism of ${\Cal S}
^{\prime}$
$$
{\cF}(\bT_{G^{(2)}(a,b;\cdot)})(-p)=e^{-p 
\beta}\cF(\bT_{G^{(2)}(b,a;\cdot)})(p)
$$
\enddemo
\remark{Remark}
In the above case the state $\omega$ is not $\cD$-continuous in the Hilbert 
space topology.
\endremark
Conversely, we show that among quasi-free gauge invariant states this is the
only posibility.
\proclaim{Theorem 3.7}
Suppose that ${\Bbb B}$ is positive, closed sesquilinear form on $\cD$ with domain
$D({\Bbb B})$. Let $\bh \geq 0$ and $\text{Ker}\, \bh ={0}$. If $e^{it\bh}\,:\,
D({\Bbb B})\to D({\Bbb B})$ and a state $\omega$ determined
by ${\Bbb B}$ by formula (4.1)
is an $\alpha_{t}$-KMS state at inverse temperature $\beta$, defined on
the Weyl algebra $\fW(D({\Bbb B}),\sigma)$, then
$$
{\Bbb B}(f,g)=\inprod{B^{1/2}f}{B^{1/2}g}\quad\text{where}\quad
B=\frac{\bI+e^{-\beta \bh}}{\bI-e^{-\beta \bh}}
$$
\endproclaim
\demo{Proof}
Let $f\in D({\Bbb B})$. Then
$$
\int\limits_{0}^{\infty}d\inprod{(B+\bI)f}{E(\lambda)f}\varphi(-\lambda)
+\int\limits_{0}^{\infty}d\inprod{(B-\bI)f}{E(\lambda)f}\varphi(\lambda)=
$$
$$
\int\limits_{0}^{\infty} d\inprod{(B+\bI)f}{E(\lambda)f}e^{-\beta \lambda}
\varphi(\lambda) +
\int\limits_{0}^{\infty} d\inprod{(B-\bI)f}{E(\lambda)f}e^{\beta \lambda}
\varphi(-\lambda)
$$
where $E(\lambda)$ are spectral projectors of $\bh$. Suppose that $\varphi
>0$ for every $x\in \bR$. Then
$$
\inprod{(B-\bI)f}{\varphi(\bh)f}=
\inprod{(B+\bI)f}{e^{-\beta \bh}\varphi(\bh)f}
$$ 
By polarization we get
$$
\varphi(\bh)(B-\bI)f=\varphi(\bh)e^{-\beta \bh}(B+\bI)f
$$
But $\varphi(\bh)$ is injective, hence
$$
(B-\bI)f=e^{-\beta \bh}(B+\bI)f
$$
So
$$
(\bI-e^{-\beta \bh})Bf=(\bI+e^{-\beta \bh})f
$$
It implies that $f\in D((\bI-e^{-\beta \bh})^{-1}(\bI+e^{-\beta \bh}))$
and
$$
\frac{\bI+e^{-\beta \bh}}{\bI-e^{-\beta \bh}}f=Bf
$$
Thus
$$
B\subset \frac{\bI+e^{-\beta \bh}}{\bI-e^{-\beta \bh}}
$$
Because both of them are self-adjoint, $D(B)=D((\bI+e^{-\beta \bh})^{-1}
(\bI-e^{-\beta \bh}))$ and
$$
B=\frac{\bI+e^{-\beta \bh}}{\bI-e^{-\beta \bh}}
$$
\enddemo
\remark{Remark}
The results of Sections 3.3 and 3.4 can also be extended to this
case. In particular, the functions $S(f,g;s)$ restricted to
$D({\Bbb B})_{+}=\{ f\in D({\Bbb B})\,:\, Cf=f \}$ define periodic Gaussian
OS-positive stochastic process indexed by $D({\Bbb B})_{+}$,
where $D({\Bbb B})_{+}$ is the real Hilbert space with respect to
the inner product
$$
\inprod{f}{g}_{B}=\inprod{B^{1/2}f}{B^{1/2}g}
$$
Similarly, since $e^{it\bh}$ is  unitary with respect to the inner
product $\inprod{\cdot}{\cdot}_{B}$, Theorem 3.5 can be applied
to this case. Moreover, the mapping
$$
f\to \omega(W_{f})
$$
is continuous with respect to the norm $||\cdot||_{B}=
\inprod{\cdot}{\cdot}_{B}^{1/2}$. Thus also in this case the process
$\xi^{\beta}_{s}$ determines the modular structure associated with
the state $\omega$. Similarly, one can show that $\xi^{\beta}_{s}$
has a version with two-sided Markov property on $K_{\beta}$.
\endremark

\subhead 3.7 Ground state process
\endsubhead
Now consider a quasi-free state $\omega_{0}$ which is a ground state
with respect to the evolution $\alpha_{t}$ defined by $T_{t}=
e^{it {\bold h}}$.
Then,
for every $f,g \in \cD$ there exists a function $\varPhi_{0}(f,g;z)$, 
analytic and bounded for $\text{Im}\, z >0$, continuous on $\text{Im}\, z 
\geq 0$ and such that$$
\varPhi_{0}(f,g;t)=F(f,g;t)
$$
for all $t \in \bR$. We know that for ground state case
$
\text{supp}\,{\Cal F}({\Bbb T}_{F}) \subseteqq [0, \infty)
$. Similar arguments as used in the proof of Theorem 3.1 lead to
the conclusion that for $\omega_{0}$,
$$
F(f,g;t)=\inprod{f}{e^{it {\bold h}}g}
$$
and ${\bold h} \geq 0$.
By analytic continuation to the Euclidean region we obtain the functions
$$
S_{0}(f,g;s)=\varPhi_{0}(f,g;is) \eqno (3.13)
$$
defined for $s\geq 0$. Two point Euclidean Green functions are now given by
$$
G_{0}^{E,2}(f,g;s)=e^{-\frac{1}{4}({\Bbb B}(f,f)+{\Bbb B}(g,g)-
\frac{1}{2}S_{0}(f,g;s)}
$$
\proclaim{Proposition 3.6}
$S_{0}(f,g;s)$ defined by (3.13) satisfy:
\roster
\item
For every $f\in \cD$ and $s\in [0,\infty)$, $S(f,f;s)$ is OS-positive.
\item
More generally, for every terminating sequences $f_{k}\in \cD, s_{k}\in
[0,\infty), c_{k}\in \bC$
$$
\sum\limits_{k,l}\conj{c}_{k}c_{l}S_{0}(f_{k},f_{l};s_{k}+s_{l}) \geq 0
$$
\endroster
\endproclaim
\demo{Proof}
Since the function 
$$
t\to F(f,f;t)=\inprod{f}{e^{it\bh}f}
$$ is positive definite, $S_{0}(f,f;s)$ is OS-positive as a Laplace
transform of positive measure. The general positivity condition
follows by polarization.
\enddemo
Similarly as in non zero temperature case, the existence of complex
conjugation $C$ defining the abelian splitting $({\cD}_{+},{\cD}_{-})$
of $\cD$, commuting with ${\bold h}$ is necessary for stochastic positivity
of $S_{0}(f,g;s)$ restricted to ${\cD}_{+}$. Adopting the arguments
from Section 3.3 to the ground state case we can also prove the following
Proposition.
\proclaim{Proposition 3.7}
Let ${\bold h}$ be $C$-real. Then functions $S_{0}(f,g;s)$ restricted to
${\cD}_{+}$ satisfy:
\roster
\item
$S_{0}(f,g;s)=S_{0}(g,f;s),\quad s\in [0,\infty)$
\item
$S_{0}(f,g;s)$ can be extended to the function of $s\in \bR$ (denoted by the 
same symbol) such that for all terminating sequences $f_{1},\ldots,f_{n}
\in {\cD}_{+}; s_{1},\ldots,s_{n}\in \bR$ and $c_{1},\ldots,c_{n}\in \bC$
$$
\sum\limits_{k,l=1}^{n}\conj{c}_{k}c_{l} S_{0}(f_{k},f_{l};s_{k}-s_{l})
\geq 0
$$
\item
$S_{0}(f,g;s)$ defines an operator valued covariance function 
$R_{\infty}(s)$ of Gaussian OS-positive stochastic process indexed by 
${\cD}_{+}$.
\endroster
\endproclaim
Let $\xi_{t}^{\infty}$ be the Gaussian process indexed by $\cD_{+}$
defined by
$$
\bE(<\xi_{t}^{\infty},f>)=0;\quad \bE(<\xi^{\infty}_{t_{1}},f_{1}>
<\xi^{\infty}_{t_{2}},f_{2}>)=
\frac{1}{2}\inprod{f_{1}}{R_{\infty}(t_{2}-t_{1})f_{2}}
$$
If in addition $\text{Ker}\, \bh =\{ 0 \}$, then as in the KMS state case,
the process $\xi^{\infty}_{t}$ corresponding
to the ground state, completely determines the ground state structure.
\proclaim{Theorem 3.8}
Let $\xi^{\infty}_{t}$ be a Gaussian, OS-positive stochastic process indexed
by $\cD_{+}\times \bR$. Then there exist a Hilbert space $\fH_{\xi}^{\infty}$
with a unit vector $\Omega_{\xi}^{\infty}$, a weakly continuous one parameter
group of unitary operators $U_{\xi}^{\infty}(t)=e^{it H_{\xi}^{\infty}}$
with $H_{\xi}^{\infty}\geq 0,\;\text{Ker}\, H^{\infty}_{\xi} = \{ 0 \}$
and a von Neumann algebra $\fM_{\xi}^{\infty}$ of operators acting on 
$\fH_{\xi}^{\infty}$ such that $\Omega_{\xi}^{\infty}$ is cyclic for
$\fM_{\xi}^{\infty}$. Moreover, the canonical ground state structure 
$(\pi_{\omega_{0}}(\WA)^{\prime \prime},e^{it 
H_{\omega_{0}}},\Omega_{\omega_{0}})$ reconstructed from the state 
$\omega_{0}$ is unitarily equivalent to the ground state structure
$(\fM_{\xi}^{\infty},e^{itH_{\xi}^{\infty}},\Omega_{\xi}^{\infty})$.
\endproclaim
\remark{Remark}
In the case $\bh > 0$, there is the following relation between covariance
$R_{\beta}$ and $R_{\infty}$. If $\beta \to \infty$, then
$$
R_{\beta}(s) \to R_{\infty}(s)
$$ 
weakly. On the other hand
$$
R_{\beta}(s)=\sum\limits_{n\in \Bbb Z} R_{\infty}(s+n\beta)
$$
where the series on the right hand side is weakly convergent.
\endremark
\head IV Examples.
\endhead
\subhead 4.1 Ground states and KMS states for quantum fields
on a stationary space-time
\endsubhead
Let $({\Cal M},g)$ be a  stationary space-time i.e. $({\Cal M},g)$
is space and time orientable with a global time-like Killing
vector field $X$.
Thus, $({\Cal M},g)$ can be always realized as $(\bR\times {\cC},g)$
where $(\cC ,\widehat{g})$ is a Riemannian 3-manifold
and
$$
g=\left(\matrix
a^{2}-b^{i}b_{i} & -b_{i}\\
-b_{i} & -\widehat{g}_{ij}\endmatrix\right)
$$
with a scalar field $a$ (laps field) and a vector field $b$ (shift field)
satisfying
$$
a >0,\quad a^{2}-\widehat{g}(b,b) >0
$$
and with the Killing vector field
$$
X:=\frac{\partial}{\partial t}=a N(\cC) +b
$$
where $N(\cC)$ is a unit future-pointing normal vector field to $\cC$.
If $({\Cal M},g)$ is globally hiperbolic, then $\cC$ can be chosen to be
a global Cauchy surface.
On $({\Cal M},g)$ we consider the covariant Klein Gordon equation
$$
(g^{\mu \nu}\nabla_{\mu}\nabla_{\nu} +m^{2}+V)\varphi=0
$$
Given some Cauchy surface $\cC$, let
$$
D(\cC)=C_{0}^{\infty}(\cC)+C_{0}^{\infty}(\cC)
$$
be the space of real smooth Cauchy data of compact support. Then, by
the Leray's theorem \cite{37}, the Cauchy data $\Phi\in D(\cC)$
given by
$$
\Phi=\left(\matrix
f\\
p\endmatrix\right)
$$
define a unique solution $\varphi$ of the Klein Gordon equation having 
compact support on every other Cauchy surface and such that
$$
\varphi |_{\cC}=f,\quad N(\cC)\varphi |_{\cC}=p
$$
Thus, we may view time evolution as a one-parameter group
$$
T_{t}\,:\, D(\cC)\to D(\cC)
$$  
Moreover, $T_{t}$ preserves the symplectic form
$$
\widehat{\sigma}(\Phi_{1},\Phi_{2})=\int\limits_{\cC}(f_{1}p_{2}-p_{1}f_{2})
\,d\eta(\cC)
$$
where $\eta(\cC)$ is the Riemannian volume element on $(\cC,\widehat{g})$
and 
$$
\frac{d}{dt} T_{t}\Phi|_{t=0}=-\widehat{h}\Phi
$$
with $\widehat{h}=-{\bold g}{\bold A}$ and
$$
{\bold A}=\left(\matrix
-(\partial^{i}a)\partial_{i} +a(m^{2}-\Delta(\cC)+V) &
-(\nabla_{i}b^{i}+b^{i}\partial_{i})\\
b^{i}\partial_{i} & a \endmatrix\right);\quad 
{\bold g}=\left(\matrix
0 & 1\\
-1 & 0 \endmatrix\right)
$$
where $\Delta(\cC)$ is the Laplace-Beltrami operator on $(\cC,\widehat{g})$
and $\nabla_{i}$ is the covariant derivative on $(\cC,\widehat{g})$.
\par
In order to apply the results of Section III we need a Hilbert space
$(\cD,\inprod{}{}_{\cD})$ containing $D(\cC)$, such that 
$\widehat{\sigma}(\cdot,\cdot)=\text{Im}\,\inprod{\cdot}{\cdot}_{\cD}$ and
$T_{t}=e^{it\bh}$, where $\bh$ is self-adjoint on $\cD$. The result of
Kay \cite{38} shows  that such {\it a one-particle Hilbert space structure}
exists under the following (mass gap) assumptions:
\roster
\item
$V$ is stationary and $\text{inf}\, V(x) +m^{2}>0$ 
\item
$\text{inf}\, a >  0$ on $\cC$,
\item
$\text{inf}\,(a-b^{i}b_{i}/a) > 0$ on $\cC$.
\endroster	
The above assumptions imply also that the generator $\bh$ has a bounded inverse.
\proclaim{Proposition 4.1}
Let $(\cD,e^{it\bh})$ be the one-particle Hilbert space structure
corresponding to the Klein Gordon equation on a globally hyperbolic
stationary spacetime $({\Cal M},g)$. There exists a complex conjugation
$C$ on $\cD$ such that $\bh$ is $C$-real.
\endproclaim
\demo{Proof}
To show the existence of complex conjugation $C$ we quote basic steps
of the Kay's construction. Since
$$
\inprod{\Phi}{{\bold A}\Phi}_{L^{2}+L^{2}} > \varepsilon ||\Phi||^{2}
_{L^{2}+L^{2}}
$$ 
for some $\varepsilon >0$, we define the ${\bold A}$-norm
$$
||\Phi||_{{\bold A}}^{2}=\inprod{\Phi}{{\bold A}\Phi}_{L^{2}+L^{2}}
$$
and $D_{{\bold A}}$ as the completion of $D(\cC)$ in this norm. Then,
it can be shown that $\widehat{h}$ is essentially skew-adjoint and 
$(\conj{\widehat{h}})^{-1}$ exists and is bounded. Since
$\widehat{\sigma}$ is continuous in ${\bold A}$-norm, it can be extended
to $\widehat{\sigma}^{\prime}$ on $D_{{\bold A}}$ and
$$
\widehat{\sigma}^{\prime}(\Phi,\Psi)=
\inprod{\Phi}{(\conj{\widehat{h}})^{-1}\Psi}_{{\bold A}}
$$
Let $D^{\bC}_{{\bold A}}$ be the natural complexification of the
real Hilbert space $D_{{\bold A}}$. We change the inner product
in $D^{\bC}_{{\bold A}}$ by modifying the complex structure on
$D^{\bC}_{{\bold A}}$. To do this we introduce a unitary operator
$I$ satisfying
$I^{2}=-1$ and $\widehat{\sigma}^{\prime}(\Phi,I\Phi)>0$. Since
$i\conj{\widehat{h}}$ is self-adjoint on $D^{\bC}_{{\bold A}}$
we can define
$$
I=i(P_{+}-P_{-})=|\conj{\widehat{h}}|^{-1} \conj{\widehat{h}}
$$
where $P_{+},P_{-}$  are  projectors onto positive and negative 
parts of its spectrum. Now we define the new inner product by
$$
\inprod{\Phi}{\Psi}_{\cD}=\widehat{\sigma}^{\prime}(\Phi,I\Psi)
+i\widehat{\sigma}^{\prime}(\Phi,\Psi)
$$
Let $\cD$ be the Hilbert space completion of $D^{\bC}_{{\bold A}}$
with respect to $\inprod{}{}_{\cD}$. From the construction of $\cD$ we see 
that $T_{t}$ is the unitary group with respect to the inner product 
$\inprod{}{}$ and the generator $\bh$ of it is strictly positive. The 
complex conjugation $C$ in $\cD$ is defined as follows. Let $C_{0}$ 
be the  complex conjugation in $D^{\bC}_{{\bold A}}$  with respect
to the complex structure given by multiplication by $i$. Then
$$
C=C_{0}(P_{+}-P_{-})
$$
is the complex conjugation in $\cD$ such that $\bh$ is $C$-real.
\enddemo
Applying the results of Section III we have:
\proclaim{Corollary 4.1}
\roster
\item
Let $\omega$ be the 
KMS state at the inverse temperature $\beta$, defined on $\WA$ as in 
Theorem 3.1 and let $\xi^{\beta}_{t}$ be the corresponding Gaussian 
Markov thermal process indexed by $\cD_{+}$ (where $\cD_{+}$ is
defined by the complex  conjugation $C$). The modular structure defined by
$\omega$ is unitarily equivalent to the modular structure reconstructed
from $\xi^{\beta}_{t}$ as in Theorem 3.5.
\item
Similarly, let $\omega_{0}$ be the ground state on 
$\WA$ and let $\xi^{\infty}_{t}$ be the corresponding ground state process. 
The ground state structure defined by $\omega_{0}$ is unitarily 
equivalent to the ground state structure reconstructed from 
the process $\xi^{\infty}_{t}$ as in Theorem 3.8.
\endroster
\endproclaim
In the case of {\it a static} space-time, we can realize $({\Cal M},g)$ as
above, but with the vector field $b=0$. Then the matrix $\bold A$ becomes
diagonal
$$
{\bold A}=\pmatrix
A& 0\\
0 & a
\endpmatrix
$$
with
$$
A=-(\partial^{i}a)\partial_{i}+a(m^{2}-\Delta(\cC)+V
$$
In this case the one-particle structure can be described in more
explicite way. As was shown by Kay \cite{38}, $\cD$ can be identified
with $L^{2}(\cC,d\eta)$ and the space $D(\cC)$ of Cauchy data is mapped
into $L^{2}(\cC,d\eta)$ by
$$
(f,p) \to Yf+iY^{\ast -1}p
$$
where
$$
Y=(\conj{a}^{1/2}\conj{A}\conj{a}^{1/2})^{-1/4} \conj{a}^{1/2}
$$
and $\conj{B}$ means closure of operator $B$. The one-particle
hamiltonian $\bh$ is now given by
$$
\bh=(\conj{a}^{1/2}\conj{A}\conj{a}^{1/2})^{1/2}
$$
and the complex conjugation $C$ is the natural one in $L^{2}(\cC,d\eta)$.
Then $\cD_{+}$ is defined as Hilbert space completion in
$L^{2}(\cC,d\eta)$ of the linear space generated by vectors
$$
Yf\,;\quad f \in C^{\infty}_{0,\bR}(\cC)
$$ 
and similarly, $\cD_{-}$
is the completion in $L^{2}(\cC,d\eta)$ of linear space
generated by 
$$
Y^{\ast -1}p\,;\quad p \in C^{\infty}_{0,\bR}(\cC)
$$
From the general results of Section III it follows that there exist
thermal (respectively ground state) processes indexed by $\cD_{+}$ and
$\cD_{-}$. The former we call {\it field
thermal} (respectively {\it ground state}) {\it process} and denote
by $\xi^{\beta}_{t}$ (respectively $\xi^{\infty}_{t}$)
and the later we 
call {\it momentum thermal} (respectively {\it ground state}) {\it 
process} and denote by $\pi^{\beta}_{t}$ (respectively 
$\pi^{\infty}_{t}$). It is enough to consider the covariances of
these processes for the indexes of the form $Yf$ or $Y^{\ast -1}p$ with
$f,p \in C^{\infty}_{0}(\cC)$. Thus for thermal processes we obtain
$$
\align
\bE(<\xi^{\beta}_{t},f><\xi^{\beta}_{s},g>)&=
\int\limits_{\cC\times \cC} (Yf)(x)R^{\beta}(t,s;x,y)(Yg)(y)
d\eta(x)\,d\eta(y)\\
\bE(<\pi^{\beta}_{t},f><\pi^{\beta}_{s},g>)&=
\int\limits_{\cC\times \cC} 
(Y^{\ast-1}f)(x)R^{\beta}(t,s;x,y)(Y^{\ast-1}g)(y)d\eta(x)\,d\eta(y)
\endalign
$$
where
$$
R^{\beta}(t,s;x,y)=
\frac{e^{-|t-s|\bh}+e^{-(\beta-|t-s|)\bh}}{1-e^{-\beta \bh}}\,(x,y)
$$
for $|t-s|\leq \beta$. Similarly, for the ground state processes
$$
\align
\bE(<\xi^{\infty}_{t},f><\xi^{\infty}_{s},g>)&=
\int\limits_{\cC\times \cC} (Yf)(x)R^{\infty}(t,s;x,y)(Yg)(y)
d\eta(x)\,d\eta(y)\\
\bE(<\pi^{\infty}_{t},f><\pi^{\infty}_{s},g>)&=
\int\limits_{\cC\times \cC} 
(Y^{\ast-1}f)(x)R^{\infty}(t,s;x,y)(Y^{\ast-1}g)(y)d\eta(x)\,d\eta(y)
\endalign
$$
where
$$
R^{\infty}(t,s;x,y)=e^{-|t-s|\bh}\,(x,y)
$$
Let $dE_{\bh}(\lambda)$ be the spectral resolution of the operator
$\bh$ in the Hilbert space $L^{2}(\cC,d\eta)$. Then
$$
\bE(<\xi^{\beta}_{t},f><\xi^{\beta}_{0},g>)=
\int\limits_{0}^{\infty}\frac{e^{-t\lambda}+e^{-(\beta-t)\lambda}}
{\lambda(1-e^{-\beta \lambda})}d\inprod{\conj{a}^{1/2}f}
{E_{\bh}(\lambda)\conj{a}^{1/2}g}
$$
Therefore, if $f$ is such that $\conj{a}^{1/2}f \in L^{2}(\cC,d\eta)$ then
$$
m(f)=\int\limits_{0}^{\infty}\frac{e^{-t\lambda}+e^{-(\beta-t)\lambda}}
{1-e^{-\beta \lambda}}d\inprod{\conj{a}^{1/2}f}
{E_{\bh}(\lambda)\conj{a}^{1/2}g} <\text{Const} ||\conj{a}||^{2}
_{L^{2}(\cC,d\eta)}
$$
and we can apply Proposition 3.6 to conclude that for every such $f$
there exists a version of the coordinate process $<\xi^{\beta}_{t},f>$
with H\"older continuous paths. Similar arguments work also for the 
ground state (field) process and momentum processes. Thus we obtain
\proclaim{Corollary 4.2}
Let $({\Cal M},g)$ be a static, globally hyperbolic space-time with the laps
field $a$. Then
\roster
\item
For any $f\in C^{\infty}_{0,\bR}(\cC)$ such that $\conj{a}^{1/2}f
\in L^{2}(\cC,d\eta)$ and $\beta \in (0,\infty]$ there exists a version of 
the coordinate process $<\xi^{\beta}_{t},f>$ such that for every
$\gamma \in (0,1/2)$ there is an integrable random variable
$d(f,\gamma)$ such that
$$
|<\xi^{\beta}_{t},f>-<\xi^{\beta}_{s},f>|\leq d(f,\gamma)\,|t-s|^{\gamma}
$$
with probability one.
\item
For any $f\in C^{\infty}_{0,\bR}(\cC)$ such that $\conj{a}^{1/2}f
\in H_{1/2}(\cC,d\eta)$ where
$$
\align
H_{1/2}(\cC,d\eta)&=\quad\text{metric completion of}\quad
C^{\infty}_{0,\bR}(\cC)\quad \text{with respect to (Sobolev -like) norm}\\
&\qquad ||f||^{2}_{1/2}=\int\limits_{\cC\times \cC}
(\bh^{1/2}f)(x)(\bh^{1/2}f)(y)d\eta(x)\,d\eta(y)
\endalign
$$
the coordinate process $<\pi^{\beta}_{t},f>$ for $\beta\in (0,\infty]$ 
has a version with H\"older continuous paths similarly as in {\rm (1)}.

\endroster
\endproclaim

\specialhead Example 4.1.1
\endspecialhead
Let ${\Cal M}_{d}$ be a flat $d$-dimensional Minkowski space-time with 
the metric tensor 
$$
(g_{\mu\nu})=\pmatrix
1&0\\
0& -\bI_{d-1}
\endpmatrix
$$
Taking 
$$
\cC=\{ (x^{0},\bold x)\in {\Cal M}_{d}\,:\, x^{0}=0 \}
$$
and $V=0$, we obtain the following covariances of thermal and ground state
(field) processes:
$$
\align
\bE(<\xi^{\beta}_{t},f><\xi^{\beta}_{s},g>)&=
\int\limits_{\bR^{d-1}}\frac{d{\bold p}}{\sqrt{{\bold 
p}^{2}+m^{2}}} \widehat{f}(\bold p)\widehat{g}(\bold p)         
\frac{e^{-|t-s|\sqrt{{\bold p}^{2}+m^{2}}}+e^{-(\beta-|t-s|)
\sqrt{{\bold p}^{2}+m^{2}}}}{1-e^{-\beta\sqrt{{\bold p}^{2}+m^{2}}}}\\
\intertext{for $|t-s|\leq \beta$ and}
\bE(<\xi^{\infty}_{t},f><\xi^{\infty_{s}},g>)&=
\int\limits_{\bR^{d-1}}\frac{d{\bold p}}{\sqrt{{\bold 
p}^{2}+m^{2}}} \widehat{f}(\bold p)\widehat{g}(\bold p)         
e^{-|t-s|\sqrt{{\bold p}^{2}+m^{2}}}
\endalign
$$
where $f,g\in C^{\infty}_{0}(\bR^{d-1})$.
And similarly for momentum processes:
$$
\align
\bE(<\pi^{\beta}_{t},f><\pi^{\beta}_{s},g>)&=
\int\limits_{\bR^{d-1}}d{\bold p}\sqrt{{\bold 
p}^{2}+m^{2}} \widehat{f}(\bold p)\widehat{g}(\bold p)         
\frac{e^{-|t-s|\sqrt{{\bold p}^{2}+m^{2}}}+e^{-(\beta-|t-s|)
\sqrt{{\bold p}^{2}+m^{2}}}}{1-e^{-\beta\sqrt{{\bold p}^{2}+m^{2}}}}\\
\bE(<\pi^{\infty}_{t},f><\pi^{\infty}_{s},g>)&=
\int\limits_{\bR^{d-1}}d{\bold p}\sqrt{{\bold 
p}^{2}+m^{2}} \widehat{f}(\bold p)\widehat{g}(\bold p)         
e^{-|t-s|\sqrt{{\bold p}^{2}+m^{2}}}
\endalign
$$
The law of the process $\xi^{\beta}_{t}$ is given by the Gaussian
random field $\mu^{\beta}_{0}$ indexed by ${\Cal S}(K_{\beta}\times 
\bR^{d-1})$ and defined by
$$
\bE^{\mu^{\beta}_{0}}(<\varphi,f>)=0\quad\text{and}\quad
\bE^{\mu^{\beta}_{0}}(<\varphi,f><\varphi,g>)=
\inprod{f}{(-\Delta_{d}^{\beta}+m^{2})^{-1}g}
$$
where $-\Delta_{d}^{\beta}$ denotes $d$-dimensional Laplace operator with
periodic boundary conditions in time direction ( $-\Delta^{\infty}_{d}$ is
defined as Friedrichs extension of $-\Delta$ in $L^{2}(\bR^{d})$). The law
of the momentum process $\pi^{\beta}_{t}$ is given by the Gaussian random
field $\nu^{\beta}_{0}$ indexed by ${\Cal S}(K_{\beta}\times \bR^{d-1})$ 
and defined by
$$
\bE^{\nu^{\beta}_{0}}(<\psi,f>)=0\quad\text{and}\quad
\bE^{\nu^{\beta}_{0}}(<\psi,f><\psi,g>)=
\inprod{f}{(-\Delta_{d-1}+m^{2})(-\Delta_{d}^{\beta}+m^{2})^{-1}g}
$$
In the present case we are able to analyse the properties of
continuity of the processes in more details.
\proclaim{Proposition 4.2}
Let $\rho \in L^{2}(\bR^{d-1})\cap C(\bR^{d-1}),\, \alpha >(d-2)/2$.
Then the field process $\xi^{\beta}_{t}$ has a version realized in
the space
$$
\bigcap_{\alpha >(d-2)/2} H_{-\alpha}^{\rho}(\bR^{d-1})
$$
where
$$
\align
H_{-\alpha}^{\rho}(\bR^{d-1})&=\quad\text{metric completion of}\quad
C^{\infty}_{0}(\bR^{d-1})\quad\text{in the norm}\\
&\qquad ||f||_{-\alpha}^{\rho}=\left[\int \frac{|\widehat{f}(p)|^{2}}
{(p^{2}+m^{2})^{\alpha}}\rho^{2}(p)\,dp\right]^{1/2}
\endalign
$$
and such that for any $\alpha > (d-1)/2$:
\roster
\item
if $\beta <\infty$, there exists a Borel measurable function
$\Theta_{\beta}\,:\, {\Cal S}^{\prime}(K_{\beta}\times \bR^{d-1})
\to \bR_{+}$ such that for all $t,s \in K_{\beta}$ such that $|t-s|<1$
$$
||\xi^{\beta}_{t}-\xi^{\beta}_{s}||_{-
\alpha}^{\rho} \leq
\Theta_{\beta}\, \left(\frac{|t-s|}{\ln (1/|t-s|)}\right)^{1/2}
$$
\item
if $\beta =\infty$, then for any $n\in \bN$ there exists a Borel
measurable function $\Theta_{n}\,:\, {\Cal S}^{\prime}(\bR^{d})
\to \bR_{+}$ such that for any $t,s\in [-n,n]$ such that $|t-s|<e^{-1}$
$$
||\xi^{\infty}_{t}-\xi^{\infty}_{s}||_{-\alpha}^{\rho}
\leq \Theta_{n}\,\left(\frac{|t-s|}{\ln (1/|t-s|)}\right)^{1/2}
$$
\endroster
In the case of momentum process $\pi^{\infty}_{t}$, let $\rho$ be as above.
Then the process $\pi^{\infty}_{t}$ is realized in the space
$$
\bigcap_{\alpha >d/2} H_{-\alpha}^{\rho}(\bR^{d-1})
$$
Moreover, for any $\alpha >(d+1)/2,\, \pi^{\beta}_{t}$
has a version with H\"older continuous (in $H_{-\alpha}^{\rho}(\bR^{d-1})$)
paths and obeys inequalities similar to {\rm (1)} and {\rm (2)}.
\endproclaim
\demo{Sketch of the proof}
Let
$$
C^{\beta}_{0}(x,y)=(-\Delta^{\beta}_{d}+m^{2})^{-1}(x,y)
$$ be the covariance of the Gaussian random field $\mu^{\beta}_{0}$.
Let $\{ \chi_{\epsilon} \}$ be the net of functions satisfying:
$\chi_{\epsilon}\in C_{0}^{\infty}(\bR^{d-1}),\, \chi_{\epsilon} >0,\,
\int_{\bR^{d-1}} \chi_{\epsilon}(x)\,dx=1$, i.e. $\chi_{\epsilon} \to
\delta$ in ${\Cal D}^{\prime}(\bR^{d-1})$. Define
$$
C^{\beta}_{0,\epsilon}=C^{\beta}_{0}\ast (\chi_{\epsilon}\otimes 
\chi_{\epsilon})$$
and let $\mu^{\beta}_{0,\epsilon}$ be the corresponding Gaussian measure
on $({\Cal S}^{\prime}(K_{\beta}\times \bR^{d-1}),{\Cal B})$, where ${\Cal 
B}$ is the Borel $\sigma$-algebra of sets in ${\Cal 
S}^{\prime}(K_{\beta}\times \bR^{d-1})$. For every $t>0$ and 
$\mu^{\beta}_{0,\epsilon}$-almost every $\varphi \in
{\Cal S}^{\prime}(K_{\beta}\times \bR^{d-1})$ we can define the map
$$
x\to X^{\epsilon}_{t}(x)=\varphi(t,x)
$$
Moreover, it can be shown that $X^{\epsilon}_{t}(x)$ takes 
values in the space $H_{-\alpha}^{\rho}(\bR^{d-1})$ \cite{39}.
Suitable modification of 
$X^{\epsilon}_{t}$ gives rise to 
$L^{2}({\Cal S}^{\prime}(K_{\beta}\times 
\bR^{d-1}),\mu^{\beta}_{0}|_{H_{-\alpha}^{\rho}(\bR^{d-1})})$ stochastic 
process and its $L^{2}$ limit is easilly seen to be the process 
$\xi^{\beta}_{t}$. Thus, there exists a version of $\xi^{\beta}_{t}$
(denoted by the same symbol) which is 
$L^{2}({\Cal S}^{\prime}(K_{\beta}\times 
\bR^{d-1}),\mu^{\beta}_{0}|_{H_{-\alpha}^{\rho}(\bR^{d-1})})$ stochastic 
process and which takes values in $H_{-\alpha}^{\rho}(\bR^{d-1})$ for
all $\alpha > (d-2)/2$. Notice that for $\alpha >(d-1)/2$ 
$$
\sup\limits_{|t-s|\leq r} \left(\bE(||\xi^{\beta}_{t}
-\xi^{\beta}_{s}||_{-\alpha}^{\rho})^{ 2}\right)^{1/2} \leq \text{Const}
|t-s|^{1/2}
$$
and
$$
\lim\limits_{t\to 
s}\bE(||\xi^{\beta}_{t}-\xi^{\beta}_{s}||_{-\alpha}^{\rho})^{2}
=0
$$
Thus we can apply the continuity criterion due to Preston and Garsia
\cite{40, 41} as formulated in Theorem 5.1 in \cite{42}. This leads to
the H\"older continuity of $\xi^{\beta}_{t}$ in $H_{-\alpha}^{\rho}
(\bR^{d-1})$ for all $\alpha >(d-1)/2$ and $\rho$ as above. Similarly 
we can prove the analogous properties of the momentum processes.
\enddemo
\remark{Remark}
For more refined continuity properties and additional references
to other continuity properties
of the process $\xi^{\infty}_{t
}$
we refer to \cite{42}. Another aspects of the process $\xi^{\beta}_{t}$
are discussed in \cite{43, 35}. It is interesting to note the essential
difference between Markov properties of the field and momentum processes.
For example, if $\beta=\infty$, then the law of $\xi^{\infty}_{t}$
fulfills so called sharp Markov property \cite{6}, while the law of
$\pi^{\infty}_{t}$ satisfies another kind of Markov property discussed
in \cite{44}.
\endremark
\specialhead Example 4.1.2
\endspecialhead
Let
$$
W_{\text{R}}=\{ (T,X)\in {\Cal M}_{2}\,:\, T^{2}-X^{2} <0,X>0 \}
$$
be the right  wedge in two-dimensional Minkowski space-time.
In the hyperbolic coordinates $(\tau, x)$ defined by
$$
T=e^{x}\sinh \tau,\quad X=e^{x} \cosh \tau
$$
the action of the Lorentz boost
$$
\Lambda(t)=\pmatrix
\cosh t& \sinh t\\
\sinh t& \cosh t
\endpmatrix
$$
becomes
$$
\Lambda(t)(\tau,x)=(\tau+t,x)
$$
Let
$$
\cC= \{ (\tau,x)\,:\, \tau=0 \} =\bR
$$
Then, on
$$
\widehat{D}_{\cC}=\{ (f,\widehat{p})\,:\, f\in
C_{0}^{\infty}(\cC),\widehat{p}=e^{x}p,p\in
C_{0}^{\infty}(\cC) \}
$$
the symplectic form
$$
\widehat{\sigma}((f,\widehat{p}),(f^{'},\widetilde{p}^{'}))
=\int\limits_{\cC} (f\widehat{p}^{'}-\widehat{p}f^{'})\,dx
$$ 
is invariant under the induced action of 
$$
\widehat{\Lambda}(t)=e^{-t{\Cal L}},\quad\text{where}\quad
{\Cal L}=\pmatrix
0&-1\\
A&0
\endpmatrix
\quad\text{and}\quad A=-\frac{\partial^{2}}{\partial x^{2}}+e^{2x}m^{2}
$$
Although in this case the mass gap conditions is not fulfilled, a
one-particle Hilbert space can also be constructed \cite{45}. $\cD$
can be identified with $L^{2}(\bR,dx)$ and $\bh=A^{1/2}$. The space
$\widehat{D}(\cC)$ of Cauchy data is mapped in $L^{2}(\bR,dx)$ by
$$
(f,\widehat{p})\to A^{-1/4}f+iA^{1/4}\widehat{p}
$$
As was shown in Section 3.6 we can define thermal state (for any
$\beta >0$) on the Weyl algebra over $D(B^{1/2})\subset L^{2}(\bR,dx)$,
where
$$
B=\frac{\bI+e^{-\beta\bh}}{\bI-e^{-\beta\bh}}
$$
Using the natural complex conjugation on $L^{2}(\bR,dx)$ we obtain
thermal processes indexed by $D(B^{1/2})_{\pm}$. On a suitable
domain in $C_{0}^{\infty}(\bR)$ we can compute the covariance of the 
field process, denoted now by $l^{\beta}_{\tau}$,  
for the indexes of the form $A^{-1/4}f,\, f\in C_{0}^{\infty}(\bR)$,
and we obtain
$$
\bE(<l^{\beta}_{\tau},f><l^{\beta}_{\tau^{\prime}},g>)=
\inprod{f}{\frac{e^{-|\tau-\tau{\prime}|A^{1/2}}+e^{-(\beta-
|\tau-\tau^{\prime}|A^{1/2}}}{A^{1/2}(\bI-e^{-\beta A^{1/2}})}g}
_{L^{2}(\bR)}
$$
Let $\varphi$ be a free scalar massive quantum field on
${\Cal M}_{2}$. It is well known that for $f$ real, the field operator
$\varphi(f)$ is essentially self-adjoint on $D^{\varphi}$ defined by
$$
D^{\varphi}=\text{lh}\{ \varphi(f_{1})\cdots \varphi(f_{n})\Omega_{0}
\,:\, f_{1},\ldots,f_{n}\in C_{0}^{\infty}({\Cal M}_{2}),n\in \bN \}
$$
where $\Omega_{0}$ is the Fock vacuum. Let for any open $G\subset {\Cal 
M}_{2}$ , let $\fN_{G}$ be the von Neumann algebra generated by
$$
e(f)=e^{i\conj{\varphi(f)}}\quad\text{with}\quad \text{supp}\,f \subset
G
$$
If $G\subset {\Cal M}_{2}$ is such that
its causal complement is non empty, then by the Reeh-Schlieder theorem,
$\Omega_{0}$ is cyclic and separating for $\fN_{G}$. In particular, we can
put $G=W_{\text{R}}$ and using the Bisognano-Wichmann theorem \cite{31} we
conclude that $(\fN_{W_{\text{R}}},\Omega_{0},\Lambda(\tau))$ forms
W$^{\ast}$-KMS system at $\beta=2\pi$.
On the other hand, for any
$f\in C_{0}^{\infty}(\bR),\,t\in \bR$, the operator $\varphi(\delta_{t}
\otimes f)$ is also essentially self-adjoint on $D^{\varphi}$. Thus we can 
define the abelian von Neumann algebra $\fA_{R}$ generated by 
$$
e^{i\conj{\varphi(\delta_{t}\otimes f)}},\quad\text{with}\quad f=\conj{f}
\quad
\text{supp}\, f\subset \{ (x,0)\,:\, x>0 \}
$$
By the direct computation we also obtain
$$
\inprod{\Omega_{0}}{e^{i\varphi(\delta_{0}\otimes f)}\Lambda(i\tau)
e^{i\varphi(\delta_{0}\otimes g)}\Omega_{0}}=
\bE(e^{i<l^{2\pi}_{0},f>}e^{i<l^{2\pi}_{\tau},g>})
$$
for $\tau \in [0, 2\pi],\,\text{supp}\,f,\,\text{supp}\,g \subset
\{ (x,0)\,:\, x>0 \}$. Moreover, it is easy to show that the modular
structure reconstructed from the thermal (boost) process $l^{2\pi}_{\tau}$
is unitarily equivalent to $(\fN_{W_{\text{R}}},\Omega_{0},\Lambda(\tau))$.
\remark{Remark}
Similar construction of the thermal process can be done in the case
of exterior Schwarzschild right wedge region in the Kruskal space-time
${\Cal M} \simeq \bR^{2}\times \Omega$ where $\Omega$ is some $d-2$-
dimensional Riemann manifold with the metric $dl_{\Omega}^{2}$. The metric
on ${\Cal M}$ is given by
$$
ds^{2}= 32 M^{3}r^{-1}e^{-r/(2M)}(dT^{2}-dX^{2})-r^{2}dl_{\Omega}^{2}
$$
where $M$ is the mass of the black hole and $r$ is the Schwarzschild radius
defined by
$$
T^{2}-X^{2}=(1-r/(2M))e^{r/(2M)}
$$
The exterior Schwarzschild right wedge is defined as
$$
{\Cal R}= \{ (T,X,\zeta)\,:\, X> |T|, r>2M,\zeta \in \Omega \}
$$
The corresponding thermal process has the covariance defined as above, but
in terms of the operator $A$ given by
$$
A=-\frac{\partial^{2}}{\partial x^{2}}+(1-2M/r)(2M/r^{3}-\Delta_{\Omega}/r^{2}
+m^{2})
$$
where $\Delta_{\Omega}$ is the Laplace operator on $\Omega$. Detailed analysis
of this process together with the construction corresponding to the de Sitter
case will be discussed in a separate paper.
\endremark
\subhead 4.2 Nonrelativistic Bose matter
\endsubhead
Let $\cD=L^{2}(\bR^{d})$ and $T_{t}=e^{it\bh}$ where  $\bh \geq 0,
\text{Ker}\,\bh =\{ 0 \}$,
commutes with the natural complex conjugation on $L^{2}(\bR^{d})$. Then
there exist thermal (and ground state) processes indexed by a suitable
subspaces of
$L^{2}_{\bR}(\bR^{d})$. In particular, then the covariance operator of
the thermal process $\xi^{\beta}_{\tau}$ is given by
$$
R^{\beta}(\tau)=\frac{e^{-|\tau| \bh}+e^{-(\beta -|\tau|)\bh}}
{\bI-e^{-\beta \bh}}
$$
for $|\tau|\leq \beta$. As in the previous examples the modular structure
defined by the canonical  KMS state at the inverse temperature $\beta$
coincides with the modular structure obtained from the thermal
process $\xi^{\beta}_{\tau}$. If additionally $\inf \sigma(\bh) >0$,
it is easy to see that for any $f\in
L^{2}_{\bR}(\bR^{d})$ the assumption of Proposition 3.4 is satisfied, hence
each coordinate process $<\xi^{\beta},f>$ has a version with
H\"older continuous paths.
\par
The case when the operator $\bh$ (called a kinetic energy) is of the form
$\bh = {\Cal E}(-i\nabla)$ where ${\Cal E}$ is some locally bounded,
mesurable function satisfying ${\Cal E}(p) \geq 0$, is of particular 
interest to physics. The choice ${\Cal E}(p)=p^{2}+\mu;\, \mu >0$
corresponds to the standard kinetic energy \cite{24}, whereas
${\Cal E}(p)=\sqrt{p^{2}+m^{2}}$ corresponds to the semirelativistic
kinetic energy. For a given ${\Cal E}\in {\Cal B}_{0}(\bR^{d})$ where
$$
\align
 {\Cal B}_{0}(\bR^{d})=&\{ \text{the set of locally bounded, 
measurable functions on $\bR^{d}$ satisfying }\\
& \inf {\Cal E}(p)=\epsilon >0 \}
\endalign
$$
we denote
$$
\widehat{C}^{\beta}(p)=\frac{1+e^{-\beta{\Cal E}(p)}}{1-e^{-\beta
{\Cal E}(p)}},\quad
\widehat{S}^{\beta}(\tau,p)=\frac{e^{-|\tau|{\Cal E}(p)}+
e^{-(\beta-|\tau|){\Cal E}(p)}}{1-e^{-\beta {\Cal E}(p)}}
$$
Since
$$
\widehat{C}^{\beta}(p)=1+\frac{2 e^{-\beta {\Cal E}(p)}}
{1-e^{-\beta {\Cal E}(p)}}
$$
the corresponding expression in $x$-space  has a $\delta$ singularity
at $x=0$. If we assume that ${\Cal E} \in C(\bR^{d})$ and
$$
{\Cal E}(p)\sim |p|^{\eta}\quad \text{for}\quad |p|\to \infty,
\eta >0
$$
we obtain
$$
C^{\beta}(x)=\delta(x)+{\Cal R}^{\beta}(x)
$$
where ${\Cal R}^{\beta}$ is continuous and
$$
{\Cal R}^{\beta}\in \bigcap\limits_{p\geq 1}L^{p}(\bR^{d})
$$
The kernel $\widehat{S}^{\beta}(\tau,p)$ has the similar properties.
\remark{Remark}
The  technical difficulties arise  only in the case when the set
$$
{\Cal N}_{\Cal E}= \{ p\in \bR^{d}\,:\, {\Cal E}(p)=0 \}
$$
is non-empty. In that case, we have to  restric the index
space of the process $\xi^{\beta}_{\tau}$. For example, if ${\Cal E}$
is continuous and 
$$
{\Cal E}(p)\sim |p|^{\nu}\quad\text{as}\quad |p|\to +0
$$
we can take as the index space $C^{\infty}_{\eta^{\ast}}(\bR^{d})$
defined as follows
$$
C^{\infty}_{\eta^{\ast}}(\bR^{d})= \{ f\in {\Cal S}(\bR^{d})\,:\,
\frac{\partial^{|\bold i|}}{\partial x^{i_{1}}_{1}\cdots 
\partial x^{i_{n}}_{n}}f\vert_{x=0}=0,\quad\text{for all}
\quad |\bold i|\leq \eta^{\ast} \}
$$
where $\eta^{\ast}\geq d-1 -\nu$. 
\endremark
Using the similar arguments as in the proof of Proposition 4.2, we
obtain the following:
\proclaim{Proposition 4.3}
\roster
\item
Let ${\Cal E}\in {\Cal B}_{0}(\bR^{d})$. Then for any $\alpha >d/2$,
there exists a version of the process $\xi^{\beta}_{\tau}$ with vaulues
in the space
$$
H_{-\alpha}(\bR^{d})=\{ f\in {\Cal S}^{\prime}(\bR^{d})\,:\,
\int_{\bR^{d}}\frac{|\widehat{f}(p)|^{2}}{(p^{2}+1)^{\alpha}}dp <
\infty \}
$$
If in addition ${\Cal E}(p)\sim |p|^{\eta}$ for $|p|\to \infty$, then
for any $\alpha >(d+\eta)/2$ there exists a version of $\xi^{\beta}_{\tau}$
such that with probability one
$$
||\xi^{\beta}_{\tau}-\xi^{\beta}_{\tau^{\prime}}||^{2}_{-\alpha}
\leq \Theta_{\beta} 
\left(\frac{|\tau-\tau^{\prime}|}{\ln(1/|\tau-\tau^{\prime}|)}\right)^{1/2}
$$
for sufficiently small $|\tau-\tau^{\prime}|$ and some integrable
random variable $\Theta_{\beta}$.
\item
Let ${\Cal E}$ be such that ${\Cal E}(p)\sim |p|^{\eta}$ as $|p|\to +0$ and
${\Cal E}(p)\sim |p|^{\nu}$ as $|p|\to \infty$. Let
$$
\align
H_{-\alpha,\eta}(\bR^{d})=& \{ \text{metric completion of}\quad 
C^{\infty}_{0}(\bR^{d})\quad\text{with the norm}\\
&||f||_{-\alpha,\eta}^{2}=\int_{\bR^{d}}\frac{|p|^{\eta}|f(p)|^{2}}
{(p^{2}+1)^{\alpha}}dp \}
\endalign
$$
Then for any $\alpha >(d+\eta)/2$ there exists a version of the process
$\xi^{\beta}_{\tau}$  with values in $H_{-\alpha,\eta}(\bR^{d})$ and for
any $\alpha>(d+\eta+\nu)/2$ there exists a version of $\xi^{\beta}_{\tau}$
with H\"older continuous paths as in {\rm (1)}.
\endroster
\endproclaim
\remark{Remark}
The case of critical (standard) Bose gas defined by the thermal state
$\omega_{\text{cr}}$ on the Weyl algebra over ${\Cal S}(\bR^{d})$ given
by
$$
\omega_{\text{cr}}(W_{f})=\exp c|\widehat{f}(0)|^{2})
\exp \left[ -\frac{1}{4}\int_{\bR^{d}} \frac{1+e^{-\beta p^{2}}}{1-e^{-\beta 
p^{2}}}|\widehat{f}(p)|^{2}dp \right]
$$
where $c>0$ is a constant and $d\geq 3$, was discussed in \cite{25}.
Here we quote only some continuity results.
\par
{\it Let $d\geq 3$ and $\xi^{\beta,\text{cr}}_{\tau}$ be the thermal process
indexed by ${\Cal S}(\bR^{d})$ with the covariance given by the density
$$
S^{\beta,\text{cr}}(\tau,p)=
c\delta(p)+\frac{e^{-|\tau|p^{2}}+e^{-(\beta-|\tau|)p^{2}}}
{1-e^{-\beta p^{2}}}
$$
For any $\alpha >d/2$ there exists a version of 
$\xi^{\beta,\text{cr}}_{\tau}$ with values in $H_{-\alpha}(\bR^{d})$
and for any $\alpha>(d+1)/2$ there exists a version of 
$\xi^{\beta,\text{cr}}_{\tau}$ with H\"older continuous paths.}
\endremark
\subhead 4.3 Quantum lattice models
\endsubhead
Let $\bL$ be a countable set with elements denoted by $\bl$.
Assume that for any $\bl\in \bL$ there is a separable Hilbert space
$\cD_{\bl}$ with an unitary group $T_{\bl,t}=e^{it\bh_{\bl}}$, where
$\bh_{\bl}$ is self-adjoint and $\sigma(\bh_{\bl})\subset [\epsilon,\infty),
\epsilon > 0$. Assume also that there is a complex
conjugation $C_{\bl}$ on $\cD_{\bl}$ such that $\bh_{\bl}$ is $C_{\bl}$-
real. Let us define
$$
\cD=\bigoplus_{\bl\in \bL}\cD_{\bl},\quad T_{t}=\bigoplus_{\bl\in\bL}
T_{\bl,t},\quad C=\bigoplus_{\bl\in \bL}C_{\bl}
$$
On $\cD$ we define a symplectic form
$$
\sigma(\varphi,\psi)=\sum\limits_{\bl\in\bL}\text{Im}\, 
\inprod{\varphi_{\bl}}{\psi_{\bl}}_{\cD_{\bl}}
$$
where $\varphi=(\varphi_{\bl})_{\bl\in\bL}\in \cD$.
Consider a state $\omega$ defined on the Weyl algebra $\WA$ by
$$
\omega(W_{\varphi})=\prod\limits_{\bl\in\bL}\exp \left[ -
\frac{1}{4}\inprod{\varphi_{\bl}}{\coth \frac{\beta}{2}\bh_{\bl}\, 
\varphi_{\bl}}_{\cD_{\bl}}\right] \eqno (4.1)
$$
The state (4.1) is $\alpha_{t}$-KMS state (for $\alpha_{t}$ defined
by $T_{t}$) at the inverse temperature $\beta$. The corresponding
thermal process indexed by $\cD_{+}=\bigoplus_{\bl\in\bL}\cD_{+,\bl}$
will be denoted by $\xi^{\beta,\bL}_{\tau}$. The process 
$\xi^{\beta,\bL}_{\tau}$ is a direct sum of the site processes 
$\xi^{\beta,\bl}_{\tau}$, indexed by $\cD_{+,\bl}$ and its covariance 
is given by
$$
\bE<\xi^{\beta,\bL}_{\tau},\varphi><\xi^{\beta,\bL}_{\tau^{\prime}},\psi>=
\sum\limits_{\bl\in\bL}\bE 
<\xi^{\beta,\bl}_{\tau},\varphi_{\bl}><\xi^{\beta,\bl}_{\tau^{\prime}},
\psi_{\bl}> \eqno (4.2)
$$
From Proposition 3.4 and equation (4.2) follows that for $\varphi
=(\varphi_{\bl})_{\bl\in\bL}$ such that
$$
m(\varphi)=\sum\limits_{\bl\in\bL} m_{\bl}(\varphi_{\bl}) <\infty
$$
where $m_{\bl}(\varphi_{\bl})$ is the moment of the measure corresponding
to the covariance of $\xi^{\beta,\bl}$ (see Proposition 3.4), there
exists a H\"older continuous version of the coordinate process
\newline
$<\xi^{\beta,\bL}_{\tau},\varphi>=\sum\limits_{\bl\in\bL}
<\xi^{\beta,\bl}_{\tau},\varphi_{\bl}>$.
\par
Let $A$ be a self-adjoint, non-negative operator on $\cD$ such that
$\bigoplus_{\bl\in\bL}\bh_{\bl}+A$ is essentially  self-adjoint on 
$D(\bigoplus_{\bl\in\bL} \bh_{\bl})$. Define
$\bh_{A}=\conj{\bigoplus_{\bl\in\bL} \bh_{\bl}+A}$. The state
$$
\omega_{A}(W_{\varphi})=\exp \left[-\frac{1}{4}\inprod{\varphi}
{\frac{\bI+e^{-\beta \bh_{A}}}{\bI-e^{-\beta \bh_{A}}}\varphi}\right]
$$
is a KMS state with respect to the evolution defined by $T^{A}_{t}
=e^{it\bh_{A}}$. If $\bh_{A}$ is $C$-real, 
there exists a thermal process $\xi^{\beta,A}_{\tau}$ indexed by 
$\cD_{+}=\bigoplus_{\bl\in\bL}\cD_{+,\bl}$, with the covariance
$$
\bE<\xi^{\beta,A}_{\tau},\varphi><\xi^{\beta,A}_{\tau^{\prime}},\psi>
=\inprod{\varphi}{\frac{e^{-|\tau-\tau^{\prime}|\bh_{A}}+
e^{-(\beta-|\tau-\tau^{\prime}|)\bh_{A}}}{\bI-e^{-\beta \bh_{A}}}\psi}
$$
The law of this process is given by a centered Gaussian random field
$\mu^{\beta}_{A}$ with covariance
$$
\bE^{\mu^{\beta}_{A}}<\varPhi,\varphi\otimes f>
<\varPhi,\psi\otimes g>=
\inprod{\varphi \otimes f}
{\left[(-\frac{d^{2}}{d\tau^{2}})_{\text{per}}+\bh_{A}\right]^{-1}\psi\otimes 
g}
$$
where $(-\frac{d^{2}}{d\tau^{2}})_{\text{per}}$ is a periodized version
of the operator $-\frac{d^{2}}{d\tau^{2}}$. Similar conclusion is true
for the ground state process.
\specialhead 4.3.1 Harmonic crystal model.
\endspecialhead
Let $\bL=\bZ^{d}$ be a lattice consisting of points in $\bR^{d}$ with
all coordinates integer. To each $\bl\in \bZ^{d}$ we associate a copy
of the complex plane $\bC$ with a natural complex conjugation
$z\to \conj{z}$ and a copy of an unitary group $u_{\bl}(t)=\exp 
\frac{i}{2}t$. By the direct sum construction we obtain the Hilbert space 
$\cD=l^{2}(\bZ^{d})$ with  the complex conjugation 
$C(z_{\bl})_{\bl\in\bZ^{d}}=(\conj{z}_{\bl})_{\bl\in\bZ^{d}}$ and the 
unitary group $T_{t}=\bigoplus_{\bl\in\bZ^{d}}u_{\bl}(t)$. Defining the 
KMS state by (4.1) and proceeding as above, we obtain a thermal process
$\xi^{\beta,\bZ^{d}}_{\tau}$ which is a direct sum of the periodized
Ornstein-Uhlenbeck processes $\xi^{\beta,\bl}_{\tau}$ with covariances
$$
\bE \xi^{\beta,\bl}_{\tau}\xi^{\beta,\bl^{\prime}}_{0}=
\frac{1}{2}\frac{e^{-|\tau|/2}+e^{-(\beta-|\tau|)/2}}{1-e^{-\beta/2}}
\delta_{\bl \bl^{\prime}}
$$
where $|\tau|\leq \beta$
\par
Let now $A$ be a self-adjoint and non-negative operator on $l^{2}(\bZ^{d})$ 
with real matrix coefficients.  
Since $A$is $C$-real, $\bh_{A}= \frac{1}{2}\bI +A$ is also $C$-real, and we 
obtain a thermal process $\xi^{\beta,A}_{\tau}$ indexed by 
$l^{2}_{\bR}(\bZ^{d})$. Its covariance is given by
$$
\bE<\xi^{\beta,A}_{\tau},(z_{\bl})><\xi^{\beta,A}_{0},(z^{\prime}_{\bl})>=
\inprod{(z_{\bl})}{\frac{e^{-|\tau|\bh_{A}}+e^{-(\beta-|\tau|)\bh_{A}}}
{\bI-e^{-\beta\bh_{A}}}(z^{\prime}_{\bl})}_{l^{2}(\bZ^{d})} \eqno (4.3)
$$
In the Gibbs state approach discussed in \cite{46, 47, 48, 49}, the process
$\xi^{\beta,A}_{\tau}$ can be obtained as a perturbation of the
process $\xi^{\beta,\bZ^{d}}_{\tau}$. The perturbation is defined as
follows. Let 
$$
d\mu_{0}^{\beta}=\bigotimes_{\bl\in\bZ^{d}}d\mu_{0}^{\beta,\bl}
$$
where $d\mu_{0}^{\beta,\bl}$ is the measure on trajectories of the
periodic Ornstein-Uhlenbeck process $\xi^{\beta,\bl}_{\tau}$. Then
$$
\align
d\mu^{\beta,A}((\omega_{\bl})_{\bl\in\bZ^{d}})=&
\lim\limits_{\Lambda\uparrow \bZ^{d}} \frac{1}{Z_{\Lambda}}
\exp\left[ -\int\limits_{-\beta/2}^{\beta/2}\sum\limits_{\bl,\bk\in \Lambda}
A_{\bl \bk}\; \omega_{\bl}(\tau)\omega_{\bk}(\tau)\,d\tau\right]
\cdot d\mu_{0}^{\beta}((\omega_{\bl})_{\bl\in \bZ^{d}})\\
Z_{\Lambda}=& \int d\mu_{0}^{\beta}((\omega_{\bl})_{\bl\in \bZ^{d}})
\exp \left[-\int\limits_{-\beta/2}^{\beta/2}\sum\limits_{\bl,\bk \in \Lambda}
A_{\bl \bk}\;\omega_{\bl}(\tau)\omega_{\bk}(\tau)\,d\tau\right]
\endalign
$$
where $\Lambda$ is a finite subset of $\bZ^{d}$ and the limit is taken in 
the sense of DLR equations (\cite{50, 51}). Clearly it defines a measure
on trajectories of the process $\xi^{\beta,A}_{\tau}$.
In fact, the set of limiting Gibbs measures can contain also some non-
translationary invariant solutions \cite{52}, which differ from
choosen above only by non-zero means. This is connected with the nontrivial
kernel of the operator $\frac{d^{2}}{d\tau^{2}}+\bh_{A}$.
Using this and
general results of Section 3.5, we  obtain the following
\proclaim{Proposition 4.4}
The modular structure of the KMS state corresponding to the harmonic
crystal model with the Fock space hamiltonian $d\Gamma (\bh_{A})$
formally given by
$$
d\Gamma (\bh_{A})= " \sum\limits_{\bl\in \bZ^{d}} (-\frac{1}{2}\frac{d^{2}}
{dx_{\bl}^{2}} +\frac{1}{2}x^{2}_{\bl}) +
\sum\limits_{\bl,\bk \in \bZ^{d}} A_{\bl \bk}\; x_{\bl}x_{\bk}"
$$
where $(A_{\bl \bk})=A \geq 0$ is
determined by the thermal process $\xi^{\beta,A}_{\tau}$ with the covariance 
{\rm (4.3)}.
\endproclaim
\remark{Remark}
The result that the modular structure of the harmonic crystal model
is stochastically determined, seems to be new. For another aspects
of this model, we refer to \cite{53}. Gibbsian perturbations of
the abelian sector were discused recently in \cite{48, 49}.
\endremark
\subhead Acknowledgements
\endsubhead
A part of this work has been done during the visits of the first (R.G.)
and the third (R.O.) authors in BiBoS Research Center
under the financial support of CIPA-CT92-4016.

\enddocument